%% file: main.tex
\documentclass[sigconf,nonacm]{acmart}
\AtBeginDocument{%
  \providecommand\BibTeX{{%
    \normalfont B\kern-0.5em{\scshape i\kern-0.25em b}\kern-0.8em\TeX}}}

\usepackage[acronym]{glossaries}
\newacronym[description={Machine Learning}]{ml}{ML}{Machine Learning}
\newacronym[description={A Machine Learning library by Google}]{tf}{TF}{Tensorflow}
\newacronym[description={The Tensorflow library for JavaScript}]   {tfjs}{TFjs}{Tensorflow.js}
\newacronym[description={A form of Recurrent Neural Network}]{lstm}{LSTM}{Long Short-Term Memory}
\newacronym[description={A Machine Learning technique which focuses on storing knowledge gained while solving one problem and applying it to a different but related problem}]{tl}{TL}{Transfer Learning}
\newacronym[description={An XML-based vector image format for defining two-dimensional graphics}]{svg}{SVG}{Scalable Vector Graphics}
\newacronym[description={A file format developed by Adobe in 1992 to present documents}]{pdf}{PDF}{Portable Document Format}

\begin{document}

\title{Magic Pen: Automatic Pen Mode Switching for Document Annotation}

\author{Kevin Desousa}
\affiliation{%
  \institution{Ontario Tech University}
  \city{Oshawa}
  \state{Ontario}
  \country{Canada}}
\email{kevin.desousa1@ontariotechu.net}

\author{Adam Bradley}
\affiliation{%
  \institution{Ontario Tech University}
  \city{Oshawa}
  \state{Ontario}
  \country{Canada}}
\email{adam.bradley@ontariotechu.ca}

\author{Nathalie Henry Riche}
\affiliation{%
  \institution{Microsoft Research}
  \city{Redmond}
  \state{Washington}
  \country{USA}}
\email{nath@microsoft.com}

\author{Ken Hinckley}
\affiliation{%
  \institution{Microsoft Research}
  \city{Redmond}
  \state{Washington}
  \country{USA}}
\email{kenh@microsoft.com}

\author{Christopher Collins}
\affiliation{%
  \institution{Ontario Tech University}
  \city{Oshawa}
  \state{Ontario}
  \country{Canada}}
\email{christopher.collins@ontariotechu.ca}

\renewcommand{\shortauthors}{Desousa et al.}

\begin{abstract}

Traditional digital pen interfaces use menu buttons to change the pen mode, which results in time and cognitive load spent on round-trip interactions and mode errors from tapping small mode selection buttons. This work presents the Magic Pen, a technique which uses machine learning to automatically switch between digital pen modes without requiring explicit mode changes. Magic Pen is driven by an LSTM model trained on pen data collected from 27 participants across two studies and uses transfer learning to iteratively tune the model towards how a specific user annotates. Error mitigation techniques using a flick gesture or on-screen tap are incorporated to correct mode errors or remove a stroke quickly. We evaluated Magic Pen in a comparative study with 18 participants, followed by iterative improvements and a deployment study with 8 participants. Magic Pen was preferred compared to a conventional menu-based approach, and transfer learning allowed for greater model predictability and stability.

\end{abstract}




\maketitle

\input{1-intro}

\input{2-relatedwork}

\input{3-datacolstudy}

\input{4-design}

\input{5-comparisonStudy.tex}
\input{5b-comparisonResults.tex}
\input{5c-comparisonDiscussion.tex}

\input{6-post_study_improvements}

\input{7-transferlearning}

\input{8-webannotationsystem}

\input{9-deploymentStudy}

\input{9b-deploymentResults.tex}
\input{9c-deploymentDiscussion.tex}

\input{10-conclusion.tex}

\bibliographystyle{ACM-Reference-Format}
 \bibliography{references}

\end{document}

%% file: 1-intro.tex
\section{Introduction}

When designing a digital pen interface, a common challenge is managing the \textit{mode problem} --- to allow for multiple modes on a stylus, systems overload the stylus' function \cite{tesler1981a, saund2003stylus}. However, the method to which the pen's mode is overloaded can be problematic, as researchers have noted that even light-weight constraints on users can distract them from their primary task \cite{arvo1999computer}.

Despite modern digital pens offering many features, such as pressure and tilt tracking, existing software has not been updated to thoroughly take advantage of these features --- with many interface design elements taken from traditional desktops. An example is using a toolbar to switch tools in applications that use pen-based annotation, such as Microsoft Word or PDF Annotator. The round trip needed to move the stylus to the top of the screen to switch modes, only to return to the main area, is one of the slowdown sources in current pen interfaces~\cite{fitzmaurice2003tracking, vogel2010direct}. Our approach to the problem is to create a means of replicating the ability to switch between tools without using elements like the toolbar, thus reducing or removing the slowdown experienced due to round trips.

Document annotation is a common use case for pens, both physical and digital. While people annotate documents in personal and idiosyncratic ways \cite{ovsiannikov1999annotation}, there are some commonalities, such as using ink to write notes and underline words and highlighters to emphasize passages. Ink, highlighter, and underline tools are commonly used and frequently interleaved during document mark-up \cite{marshall1997annotation}, such as an editorial pass or a student studying for a test. The round trip for mode-switching using a menu can slow down the flow of the annotation process \cite{fitzmaurice2003tracking}.

We started our investigation of the possibilities of automatic mode switching with a data collection study (Section 3). Informed by observations and analysis of the collected data, this paper proposes (Section 4) and tests (Section 5) two competing approaches to automatically change a digital pen's mode without direct user interaction. To mitigate the impact when automatic approaches make a classification mistake, we propose a technique that allows users to change mode or remove strokes quickly with flick and tap gestures.  Both classification approaches, Machine Learning and Rule-based decision trees, take advantage of some of the commonalities in how people use these tools in document annotation, such as stroke shape and location. We study how these methods perform against a traditional menu-based approach and investigate whether the ability to learn from nuanced actions in an ML pen mode switcher makes up for the lower predictability of its execution. We then follow up on the ML approach through iterative refinement of the techniques (Section 6) and the incorporation of \acrlong{tl} to customize the classifier to individual writing styles (Section 7). The improved technique was embedded in a web-based PDF annotation tool (Section 8) which we evaluated through a 5-session deployment study (Section 9).

%% file: 2-relatedwork.tex
\section{Related Work}
There is much untapped potential for using input methods beyond a traditional mouse and keyboard to interact with a computer system~\cite{millen1993pen}. The usability and usefulness of pen interfaces are crucial in the appeal of lightweight, hand-held computers, as their small size prohibits the inclusion of typical input methods~\cite{frankish1995recognition}. In this section, we discuss past approaches to the design of automated or rapid mode-switching techniques for digital pens.

\subsection{Studies of Annotations}

A considerable amount of work investigated ink annotations to derive implications for digital interfaces. The most relevant studies on annotations are Marshall's study~\cite{marshall1997annotation} on annotations of physical textbooks and Ovsiannikov et al. \cite{ovsiannikov1999annotation} on the main types of annotations people perform and their uses. Results of these studies indicate that annotations are useful to remember, think, clarify, and share portions of text they found particularly meaningful, a critical activity identified as \textit{active reading}. Ovsiannikov et al. note that, in order for electronic annotations to be more appealing than their paper counterparts, they must be superior in the capabilities they offer, such as the ability to provide an infinite number of pen tools. Most common annotations are to mark up portions of text with a marker, followed by highlighting keywords and writing in the margins. Therefore, we focus on these types of annotations in this work.

Marshall's study investigated marks made on physical textbooks and derived implications for the design of digital annotation tools. A key insight from this work is that most annotations consists of marks directly overlaid over content such as underlines and circles, which calls for the ability to directly mark up digital content. They also noted the idiosyncratic nature of annotations and advocated for fluid tools able to support varied personal practices. Lastly, they emphasized the importance of limiting disruptions to preserve cognitive resources devoted to reading. Our work with Magic Pen focuses on providing multiple types of pen modes to offer rich annotation capabilities to readers (e.g., different strokes for highlighting, underlining or writing notes over content) while limiting disruptions caused by mode switching by relying on automatic methods.

\subsection{The Mode Problem and Optimization Efforts for Pen Interfaces}

In sketching software, special consideration must be given towards managing the \textit{mode problem} \cite{tesler1981a}. This problem mainly stems from the fact that there is an excess of commands available compared to what a digital pen can support at any given moment. Thus, most software relies on some deliberate action to select amongst the overloaded functions of the pen. In the past, actions such as pressing a button \cite{lin2000denim}, double-tapping the stylus \cite{song2011grips}, and explicit stroke gestures~\cite{ramachandran2003architecture, scriboli-hinckley2005} have been used to signify to the computer that a user wishes to change the pen's operation. However, the issue with these actions is that a user must remember to perform them before switching pen modes. If not, users find themselves in a situation where they create meaningless strokes without any indication of what they did wrong \cite{saund2003stylus}. To recover from this state, users must divert their attention from their task to remove any unneeded strokes and enter their intended mode to repeat the mistaken action.

As a result of the mode problem with current pen-interfaces, there have been efforts to optimize the mode-switching behavior in traditional desktop interfaces, including the use of external actions to invoke a particular mode. For example, DENIM~\cite{lin2000denim} and  Flatland~\cite{mynatt1999flatland} rely on pen barrel buttons that users must press or hold to switch action options (called ``pen modes.'') Mode switches using these approaches can be permanent or last only while the button is depressed (``quasi modes''). However, as noted by Saund and Lank~\cite{saund2003stylus}, this approach suffers from two critical drawbacks. Physically pressing/holding a button while making a gesture can be awkward. It also requires the users to remember to perform this initial action before switching pen modes, which may not reduce cognitive load compared to menus.%

Another approach that has been experimented with is the use of features offered by digital pens, such as pressure or tilt, to change pen modes. Xin et al. \cite{xin2012natural} investigated the natural use of pen tip pressure, tilt, and azimuth (PTA) as a means to find correlations between the different modalities that could be used to switch pen modes. Their findings conclude that the use of pressure as an explicit mode-switch signal is the most likely to cause incorrect mode-switches because of the wide range of pressure (61\% of total pressure sensitivity range) used when writing. In comparison, azimuth is the least likely due to it only using 24\% of its full range. The authors also note that PTA profiles can directly benefit adaptive mode switching, which is the main focus of our work.

Li et al. \cite{li2005experimental} extend the idea of using pressure to switch modes by investigating the performance of pressure-based mode switching, along with four other techniques. After a performance evaluation, they noted that the pressure technique, while promising, did not perform as well as the non-preferred hand method. However, they also noted that each participant had different pressure readings while using the method --- the method's performance could be significantly improved if they accommodated that personalized pressure space.

Other approaches to optimize mode-switching display a list of pen modes (highlight, underline, etc.) after a detected pen stroke. Recording and monitoring the position of the pen with respect to on-screen elements~\cite{lui2004adaptive, lui2003system} enables detection of specific gestures (delimiters) in a pen-stroke~\cite{scriboli-hinckley2005}. Systems such as Scriboli by Hinckley et al.~\cite{scriboli-hinckley2005} rely on such gestures (drawing a pig tail) to change mode. Research shows that these methods are a viable alternative to the standard menu because they are fast, unambiguous, and expressive. Yet they require users to memorize a gesture and interweave it with regular annotation.

In contrast to these previous interactions techniques, several methods attempted to infer pen modes ideally preserving the user from switching modes. The Inferred-Mode protocol, developed by Saund and Lank~\cite{saund2003stylus},
uses a decision tree to determine which mode the user intends (draw or select), based on stroke shape and context (location), displaying a menu as a fallback. A key drawback of this method is the detection of the intended mode \textit{a posteriori}, preventing the system to switch pen modes while strokes are laid out on the screen. Changing modes modifies the appearance of strokes after they are made, resulting in potentially visually disruptive experiences.

\begin{figure*}[t]
        \frame{\includegraphics[width=\textwidth]{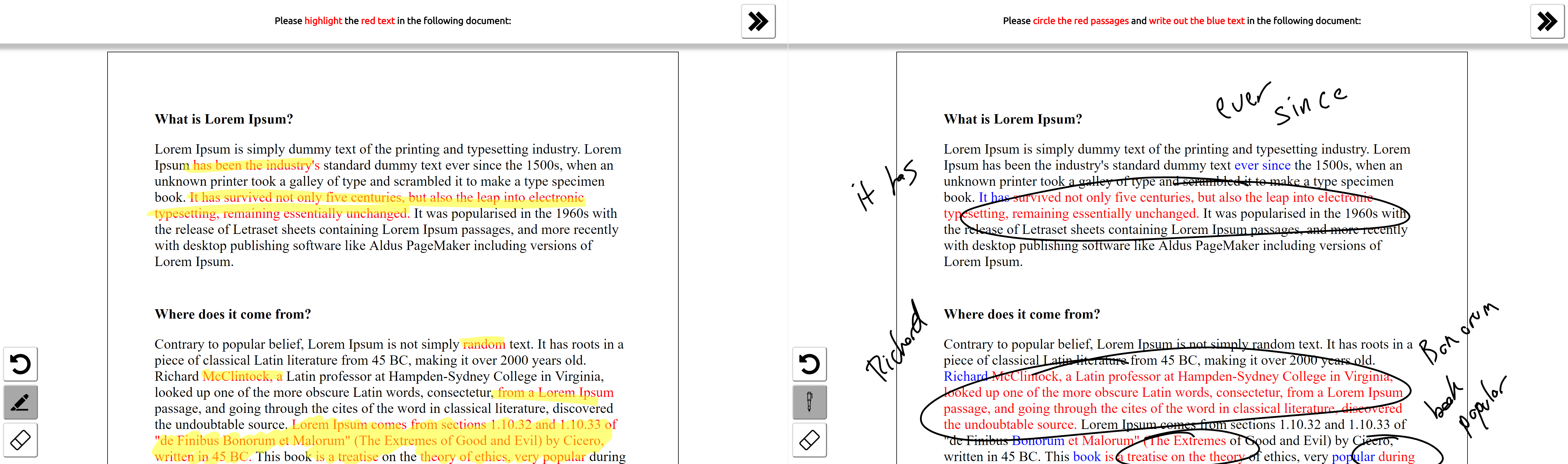}}
        \vspace{-.6cm}
        \caption[Examples of tasks in the Stroke Recorder System data collection study.]{Example task screens from the Stroke Recorder System: (left) highlight (right) draw.}

        \label{fig:stroke-recorder}
\end{figure*}

We revisit and extend the work on inferred modes through two low latency approaches for mode inference between drawing modes: deep learning and decision trees. Rather than a button for ambiguous modes, we provide a gesture-based fallback method to cycle recent strokes between modes. Analysis of a stroke's mode is done tens of times a second while a stroke is being drawn, which allows for low latency mode-detection that can be used to update the visual representation of a stroke shortly after drawing begins.

%% file: 3-datacolstudy.tex
\section{Data Collection Study}
The goal of our \textit{Magic Pen} technique is low latency automatic mode inference during document annotation. To create classifiers, we designed and ran a data collection study, approved by our ethics board, on how people hold and use a digital pen when annotating with different modes. The study collected pen-data features such as pressure and tilt, and our analysis investigated variances in these features across the modes.

\subsection{Methodology}

\subsubsection*{Participants:} We recruited 9 right-handed participants (5 male, 4 female; ages 18--26) from our university student population. Participants were screened to have experience with digital pens and unimpeded use of their right hand and wrist. We restricted participation to right-handed people to improve the likelihood of discovering consistent mode-signifying patterns, as left-handed people are known to make strokes in a different manner~\cite{gregory1980effects}. Each session lasted one hour and participants were compensated \$20 for their time.

\subsubsection*{Setup:} The study was conducted within a private, quiet lab with only one investigator present. A Microsoft Surface Pro 6 and a Surface Pen were provided to the participants to perform the study. To control external variables such as device orientation, the tablet was laid flat and taped to the table roughly 30cm away from the participant. While this restriction incurred a loss of external validity, and some participants mentioned the orientation felt unnatural at first, we wanted to collect consistent pen pressure and orientation data for a proof-of-concept which could be generalized to varying tablet setups if shown to be promising in the restricted case.

\subsubsection*{Procedure:} The study consisted of having the participants perform a series of text annotation tasks with our data collection interface which recorded each interaction. Before conducting the study, participants were  encouraged to write naturally, and reassured that they were not going to be assessed based on performance or legibility of their handwriting.

\subsubsection*{Tasks:} Participants used different pen modes to conduct 4 annotation tasks on randomly selected passages of 10--30 characters. Each task was repeated twice in counterbalanced order for a total of 4 tasks x 2 trials = 8 task instances:
\begin{description}
    \item[T1] \textbf{Highlighting} red passages from a document (Figure \ref{fig:stroke-recorder} left)
    \item[T2] \textbf{Underlining} red passages from a document, by drawing a line at the base of the text
    \item[T3] \textbf{Striking out} red passages from a document, by drawing a line through the middle of the text
    \item[T4] \textbf{Drawing} circles, brackets, stars, asterisks, and arrows near the red passages, while \textbf{writing out} text in blue (Figure \ref{fig:stroke-recorder} right)
\end{description}

Each task screen allowed the participant to switch between the task's pen tool and the eraser. An undo button was also provided. The tasks for the participants to carry out were chosen to replicate the typical actions undertaken in document annotation scenarios, such as editing a manuscript, studying, or grading papers. The draw mode, sometimes referred to as `inking', needed special consideration, as there are a variety of different types of marks one can write within that mode. We attempted to cover these possibilities within the draw mode by having the participants draw various symbols and write out text. Thus, this task consisted of several subtasks (drawing circles, brackets, stars, asterisks, arrows) which were presented in sequential screens.

After the 8 mode-specific task instances, participants were given a concluding open-ended task in which they could use all the pen tools, in any order, to markup a provided document. The only restriction was to use each of the provided tools at least once. This phase was used to collect data on stroke behaviour in more natural annotation flows, where tools can be interleaved.

\subsection{Stroke Recorder System}
To allow us to capture data about how people hold and use a digital pen, we created a web-based interface with an emphasis on three key factors:

\begin{description}
    \item[Low Latency:] User engagement with pen interfaces is directly tied to interface performance~\cite{scriboli-hinckley2005}.
\begin{figure*}[t]
    \centering
    \hspace*{-1cm}
    \includegraphics[width=\textwidth]{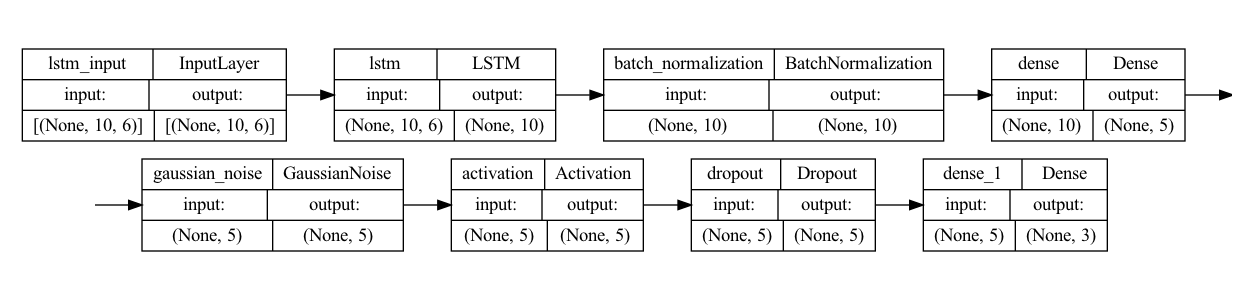}
    \caption[\acrshort{ml} Network Diagram]{The LSTM model architecture used in the \acrshort{ml} technique to determine pen modes.}
    \label{fig:mldiagram}
\end{figure*}

\begin{figure*}[t]
    \centering
    \includegraphics[width=\textwidth]{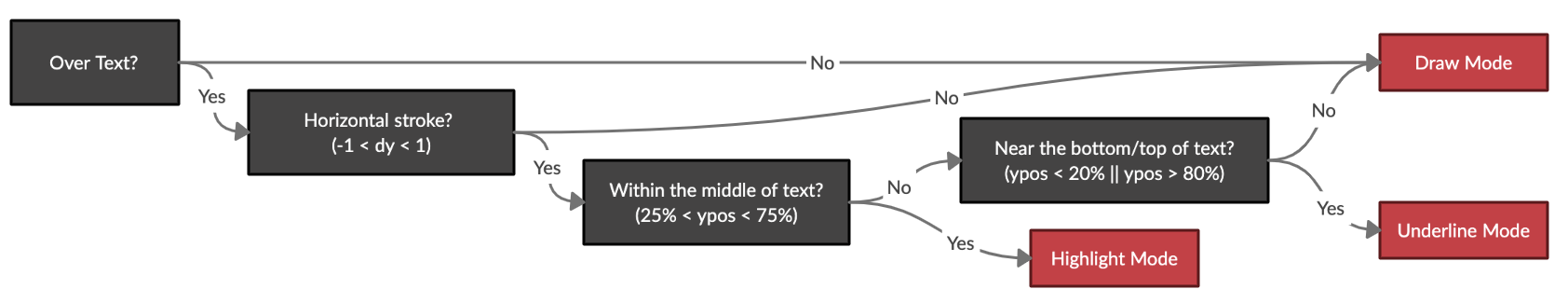}
    \caption{The decision tree used in the rule-based technique to determine pen modes.}
    \label{fig:decisiontree}
\end{figure*}

\begin{figure}[t]
        \frame{\includegraphics[width=\columnwidth]{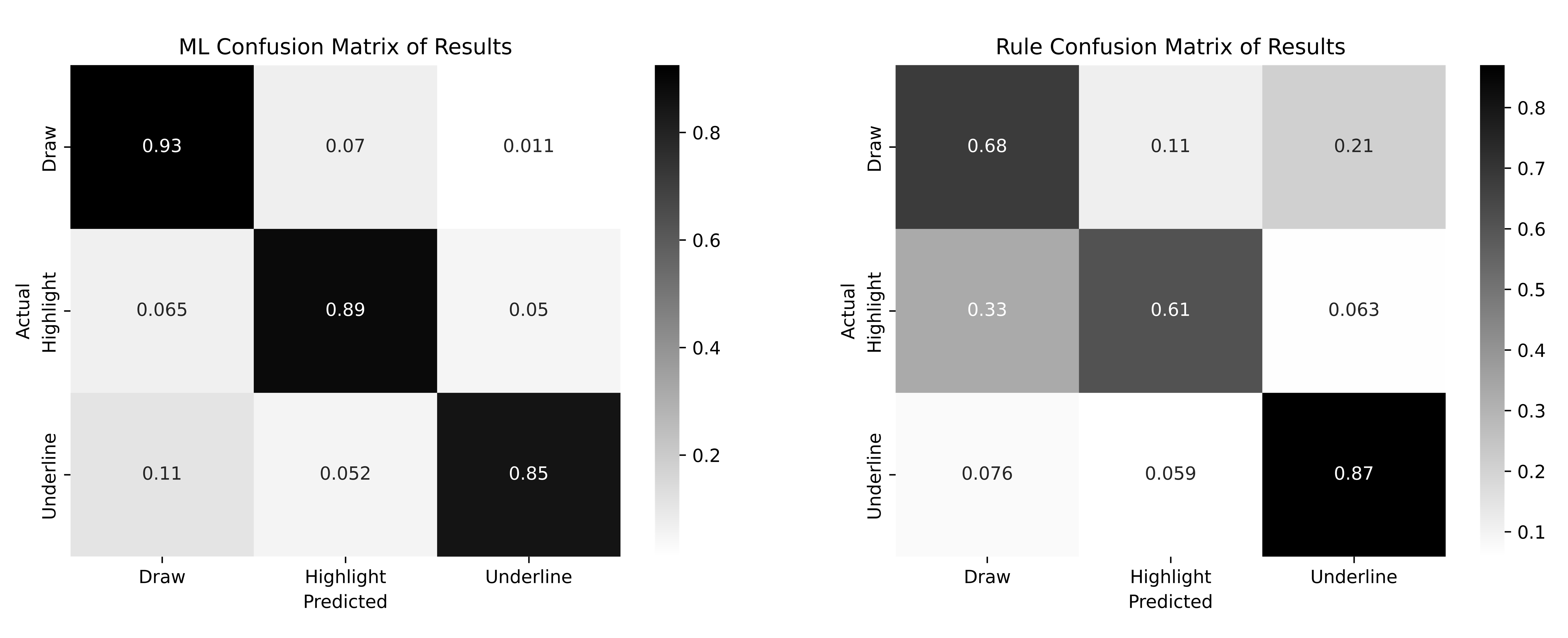}}
        \caption[Confusion matrices of both techniques.]{Confusion matrices of both techniques. Machine learning generally performs better than the Rule-based counterpart.}
        \vspace{-2mm}
        \label{fig:confmatrix}
\end{figure}

\begin{figure*}[t]
        \includegraphics[width=\textwidth]{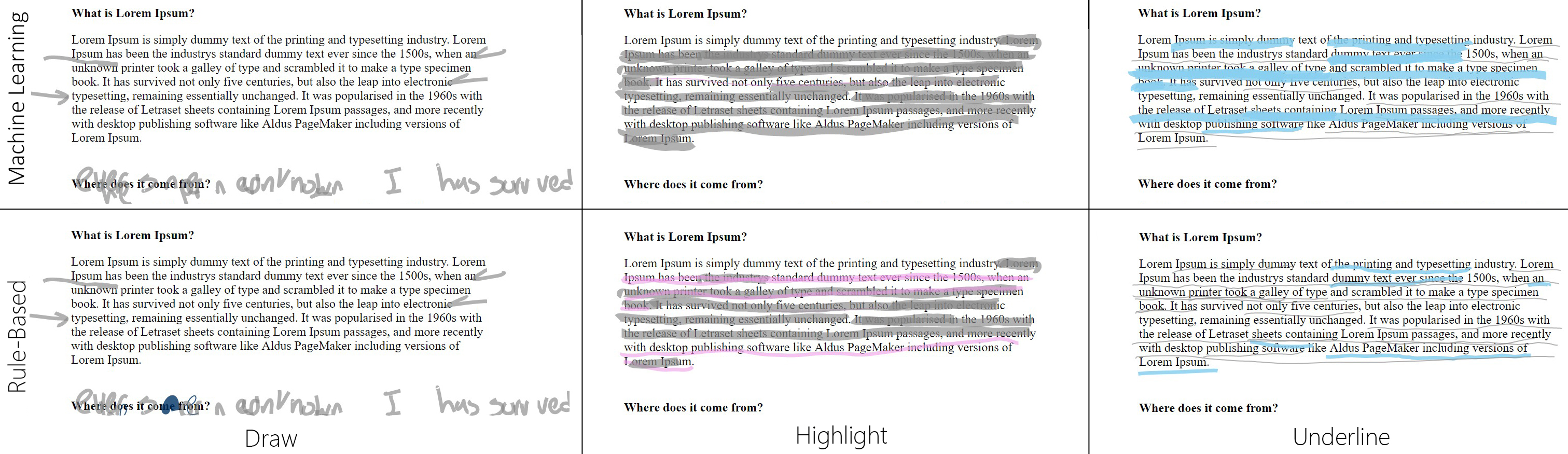}
        \caption[A visualization of errors made with the deep learning approach.]{%
         A visualization of the errors made with draw, highlight, and underline strokes (l-r) using the ML approach (top) and Rule-based approach (bottom) on the same sample of strokes. Grey strokes were classified correctly. For errors, colours (dark blue=draw, purple=highlight, and light blue=underline) represent the correct stroke class, while the stroke width (medium=draw, thick=highlight, thin=underline) represents the predicted class.
        }
        \vspace{-1mm}
        \label{fig:errorvis}
\end{figure*}

\begin{figure*}[t]
    \includegraphics[width=\textwidth]{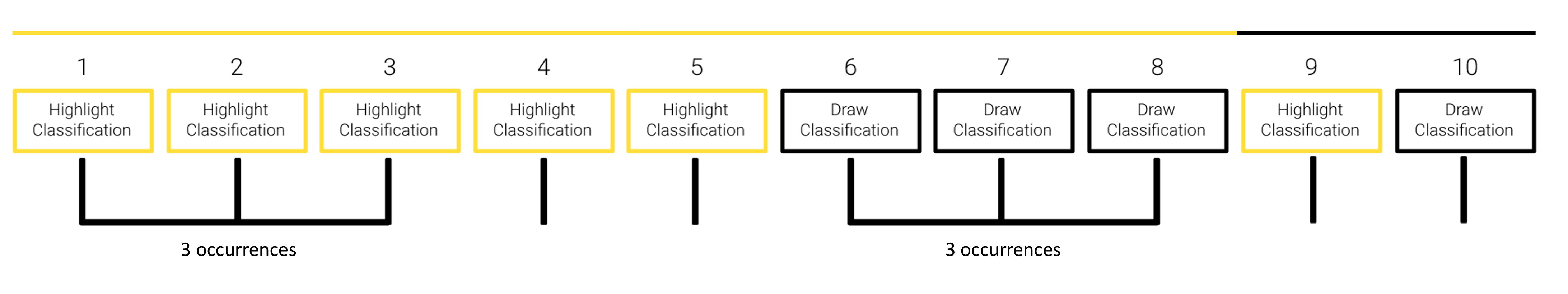}
    \caption[A Visual Representation of the Mode Selection Technique]{An example of the mode selection technique. Each block is a \textit{stroke window}.  For a mode switch to be successful, there must be 3 classifications of the same mode in a row.}
    \label{fig:FirstSeqModeSel}
    \centering
\end{figure*}

\begin{figure*}[t]
    \centering
    \includegraphics[width=\textwidth]{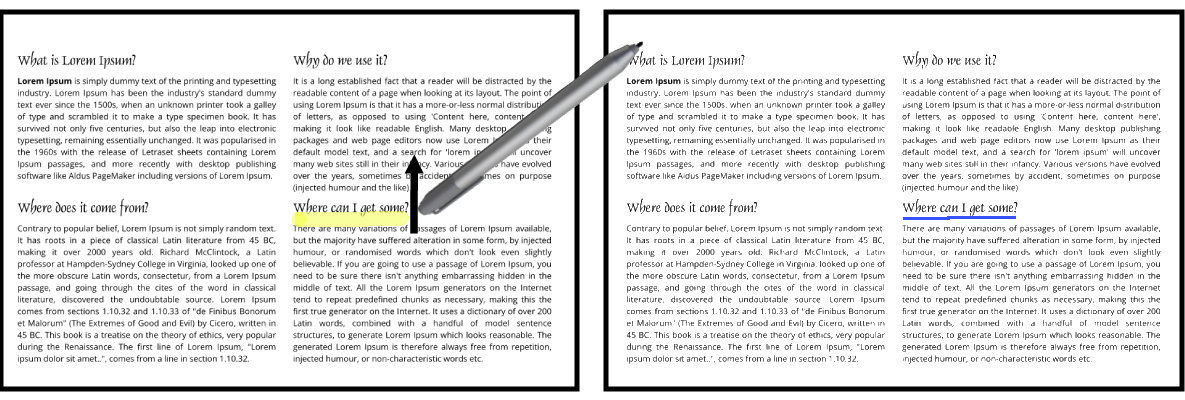}
    \caption[A Storyboard Sketch Visualizing the Eraser Mitigation Technique]{An illustration of the error mitigation technique. A flick of the eraser changes the incorrect mode from highlight to underline.}
    \label{fig:mitigationex}
\end{figure*}

    \item[Flexibility:] Support a variety of different pen tools and hardware platforms for remote deployment.
    \item[Data Completeness:] Collecting all information regarding the pen and stroke may inform the classification.
\end{description}

The software runs on any pen-enabled device compliant with the W3 pen standard~\cite{w3c-pointer}.

Using the JavaScript Pointer-Event API~\cite{pointer-event}, the stroke recorder system collected all available features relating to the digital pen, and the status of the software (\texttt{time}, \texttt{pen mode}, \texttt{text elements under stroke}).  Some API-supported features were dropped because they were stroke invariant (e.g., \texttt{device operating system}) or not supported by our hardware/browser combination (e.g., \texttt{twist}). Features were collected irregularly due to JavaScript's event-based nature, with most samples being around 10--15ms apart from each other.

\subsection{Results}
The data collection study gathered 11,338 strokes using the stroke recorder system.
As a single time steps do not contain enough data for mode classification, the raw data was pre-processed  into collections of ten stroke samples called \textit{stroke windows}.  Pen location and pressure on screen were transformed into deltas to prevent the stroke recognition model from learning absolute aspects of the tablet used within the study.

We aimed to narrow the feature set to those which carry useful information for mode inference. In addition to charting the data to observe feature variances across modes, we ran a series of experiments to determine if the ML system's performance improved with each feature's inclusion. A model was built up one feature at a time until optimal classifier performance was achieved. During these experiments, the classification of strike out vs. highlight was error-prone. Thus strike out was dropped from all further work.

\subsection{Final Feature Set}

We selected the following subset of the collected and derived features which were appeared promising through our analysis as well as informative to our LSTM classifier:

\begin{description}
    \item[\texttt{Delta X, Delta Y}:] The change in the pen's position on-screen since the last recorded sample.
    \item[\texttt{Pen Pressure}:] The pressure applied to the screen using the pen. Values range from 0.0 to 1.0.
    \item[\texttt{Delta Pen Pressure}:] The change in pressure applied to the screen since the last sample.
    \item[\texttt{Nearest Objects}:] A binary value encoding whether the pen is directly over text at any given moment.
    \item[\texttt{Closest Bottom}:] The distance to the closest bottom of a text object's bounding box. Distances are encoded as a percent of the object’s height to be font size invariant.
\end{description}

%% file: 4-design.tex
\section{Design of Magic Pen}

\input{4a-design_mlmodel}

\input{4b-design_rulemodel}

\input{4c-design_performance}

\input{4d-design_systemdesign}

\input{4e-design_mitigation}

%% file: 4a-design_mlmodel.tex
Magic Pen was initially designed with two different back-end implementations: the first uses machine learning to classify stroke modes based on data collected in our study; the second uses a decision tree informed by observations in the data collection study. The motivation for trying two approaches was that the ML approach may discover hidden feature patterns to permit accurate classification, while a decision tree classifier may be more predictable and explainable.

\subsection{Machine Learning Approach}

After collecting pen data and determining which features were most likely relevant for detecting pen modes, a stroke recognition model was trained to identify the distinct modes. To generate the model, a training script was created in Python using the TensorFlow framework~\cite{abadi2016tensorflow}. A \acrfull{lstm} network was chosen due to past success on temporal data such as stroke and speech recognition~\cite{graves2013hybrid, su2016novel}.  The model's architecture is shown in Figure \ref{fig:mldiagram}.

In general, \acrshort{lstm} layers can learn more effectively with a higher number of hidden units, but this comes at the cost of requiring a significant amount of training data to generalize. Special consideration was taken with the number of hidden units for the \acrshort{lstm} layer to allow the model to learn complex trends while not overfitting. A series of optimization tests were conducted to determine the optimal number of hidden units and hyper-parameters by iteratively running the model with different combinations of varying values (a brute-force approach). A single \acrshort{lstm} layer with ten hidden units was the best compromise between generalization and overfitting. Class weights were also calculated to even out the imbalanced training data so that the model did not overfit towards one classification. Two sets of dense layers then followed. Batch normalization and Gaussian noise regularization methods were used to avoid overfitting, as they have been used successfully in previous research~\cite{ioffe2015batch, srivastava2014dropout}. A single dropout layer was used following the activation layer for similar reasons. Layer hyperparameter values of 0.1 (Gaussian noise) and 0.15 (dropout) produced the best results.
For training the \acrshort{lstm}, a batch size of 64 and a maximum of 500 epochs was used.
The \texttt{EarlyStopping} \acrshort{tf} callback with a \texttt{patience} value of 20 \cite{abadi2016tensorflow} was used to avoid  overfitting while still allowing the model to grow.

%% file: 4b-design_rulemodel.tex
\subsection{Rule-based Approach}
When using the ML model, we observed that the features involving the pen's position on the screen with respect to other elements were strongly predictive of pen mode. For example, highlight mode classification was likely when the pen was near the middle of a word. An alternative method that didn't rely on ML seemed feasible.  Saund and Lank~\cite{saund2003stylus} made a similar discovery in creating the ``Inferred-Mode'' system --- signifying that objects on screen might be a strong-enough feature to switch modes. We developed an alternative Rule-based approach. While simpler, this method has the drawback of not leveraging features such as pressure or motion dynamics to differentiate stroke types that occur outside the presence of text elements (e.g., in the margins).  We chose a manually constructed decision tree to implement the rule-based mode-switching system. The decision tree rules were selected by visualizing the initial study's resultant stroke data and observing the relative location over text for each pen mode. The resulting decision tree is seen in Figure~\ref{fig:decisiontree}. For consistent system architecture with the ML approach, the decision tree also uses stroke windows to make a classification.

%% file: 4c-design_performance.tex
\subsection{Model Performance}
\begin{figure*}[t]
    \includegraphics[width=0.56\textwidth]{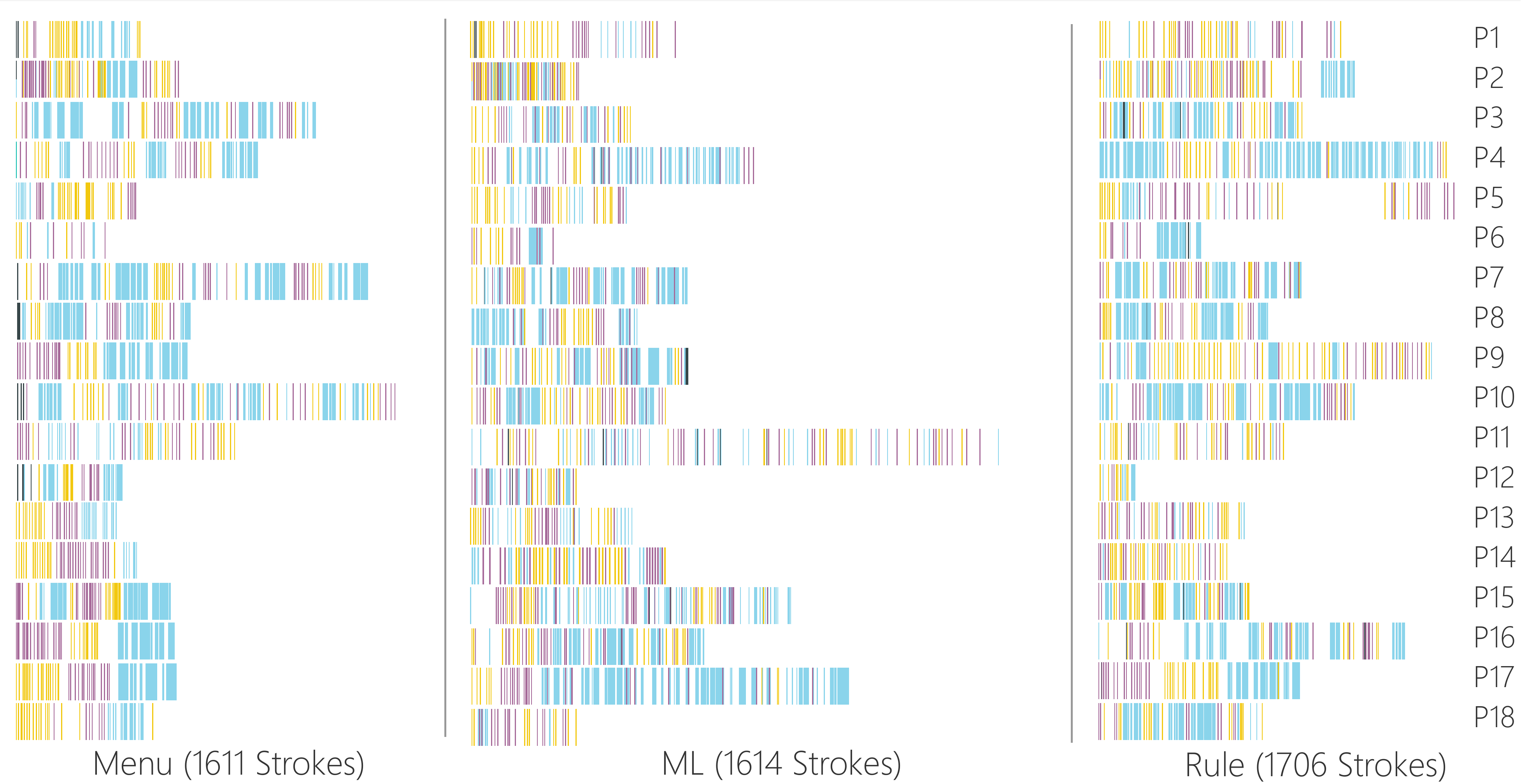} \quad \includegraphics[width=0.40\textwidth]{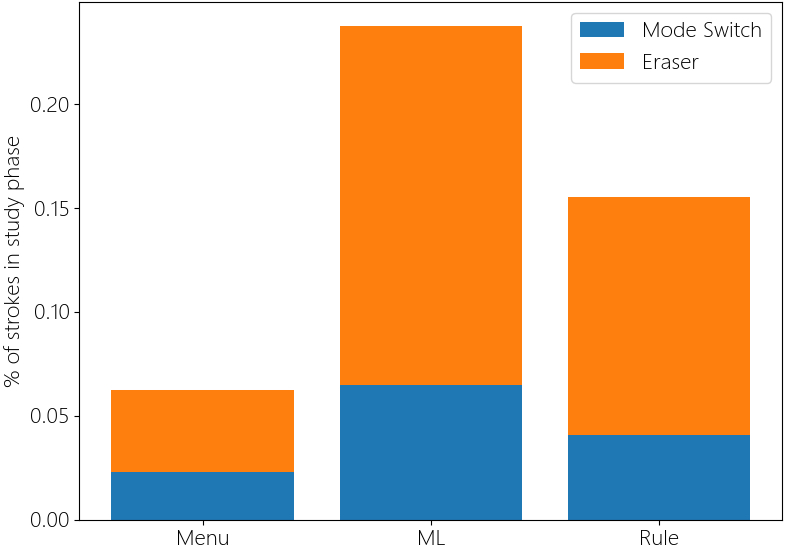}

    \caption{(Left) Temporal stroke sequences made by each participant, binned for each study condition. Blue represents draw, purple is underline, yellow is highlight, and grey is unclassified (too short to classify or no mode selected). (Right) The percent of edited (blue) and erased (orange) strokes in each study condition by the technique used.}%
    \label{fig:strokeresults}
\end{figure*}

We assessed each method's performance by performing K-Fold Cross Validation with 10 folds, resulting in a set of 11,338 strokes randomly sampled from the 9 participants. After training on 293,997 stroke windows, the \acrshort{ml} network achieved an overall accuracy of 88.55\%, while the Rule-based approach achieved a 72.67\% accuracy (see Figure \ref{fig:confmatrix}).

\begin{table*}[t]
    \centering
    \begin{tabular}{@{} rcl }
         \toprule
         \textbf{\#} & \textbf{Question Criteria} & \textbf{Question} \\
         \midrule
         1 & None & The technique switched to the correct pen modes I attempted to use. \\
         2 & None & I felt efficient with the technique. \\
         3 & None & I felt frustrated with the \textit{<colour>} technique. \\
         4 & None & I felt distracted by the \textit{<colour>} technique. \\
         5 & None & I felt the \textit{<colour>} technique required more effort than usual. \\
         6 & None & I felt the result of the \textit{<colour>} technique was predictable. \\
         7 & None & I felt I adapted my interaction with the pen as I used the \textit{<colour>} technique. \\
         \midrule
         8 & Used Eraser Flick & I found the ``flick'' eraser gesture to be a fast way to correct the mistake. \\
         9 & Used Eraser Flick & I found the ``flick'' eraser gesture to be an intuitive way to correct the mistake. \\
         \midrule
         10 & Did Not Use Eraser Flick & Did you have any errors that you needed to correct? \\
         \bottomrule
    \end{tabular}
    \caption[Evaluation Study Questionnaire Questions]{
        Feedback questions and the criteria for each question being displayed.  Techniques were represented by colours in the study --- red, blue, or green. Answers for Q1--9 were on a 5-point Likert scale from \textit{strongly disagree} to \textit{strongly agree}, and Q10 was yes or no.
    }
    \label{tab:EvalStudy1Q}
\end{table*}

As shown in Figure \ref{fig:errorvis} (top), the main benefit to using ML is that it can detect highlight and draw  consistently, despite the users making strokes that appear to be at the top of the words (in the case of highlight) or even drawing over text. The majority of errors were with the underline mode. Figure \ref{fig:errorvis} (bottom) shows that the Rule approach can determine draw strokes consistently but makes errors between the other two modes. However, an interesting product of this technique's classification is that the results are reasonably predictable between modes. This predictability can be seen in the underline strokes where the Rule approach misclassified because the stroke at points was not near the top or bottom  of the text. The same can be seen with the highlight strokes, where the classification became draw for the same reason. These errors may be mitigated by users adapting their writing to the model, though this goes against our goal of naturalistic automatic classification.

%% file: 4d-design_systemdesign.tex
\subsection{Real-time Mode Selection}

Given a stroke \textit{sample} as an input, the mode-switching system begins with a pre-processing step, to determine whether there are enough (10) time samples to create a \textit{window}. If not, it then checks to see whether the stroke ended and if there were at least two stroke samples recorded. If at least two samples were collected, linear interpolation is used to generate the remaining samples to create a window.

Once time samples have been grouped into a window, a \textit{classification} is generated using either the ML or Rule techniques. Each classification is stored in an array for that \textit{stroke} (each time a window of the stroke is classified, extending the array). The displayed stroke mode is then determined by getting the $N$ most recent consecutive classifications from the same class to determine the final pen mode. An $N$ of 3 was found to be an optimal balance of incorporating history but preferring recency (see Figure~\ref{fig:FirstSeqModeSel}).

%% file: 4e-design_mitigation.tex
\subsection{Mitigating Errors}
\label{chap:EraserMitigation}

Incorrect classifications can frustrate users, which may lead them to avoid using a mode-switching technique altogether~\cite{vogel2010direct}. We created an error-mitigation technique that avoids traditional toolbar menus by using the eraser-end of the pen to perform a flick gesture. Initially, pen tip + button was considered for this task, as it would be quick. However, we settled on using the pen's eraser end to give a clear distinction between stroke-correction and inking. The correction gesture is simply a flick up or down on an empty part of the screen to cycle the last-drawn stroke's mode (Highlight $\rightarrow$ Underline $\rightarrow$ Draw) (see Figure~\ref{fig:mitigationex}). The eraser on the pen acts as a regular eraser when moving over an existing stroke. Barrel button-activated erasers can also be used with the flick gesture. If no dedicated eraser is provided, a modal toolbar button was added to activate eraser mode.

%% file: 5-comparisonStudy.tex
\section{Magic Pen Comparison Study}

We conducted a remote-deployment mixed-methods evaluation study to compare our two automatic mode switching methods against each other and traditional menu-driven modes. The study collected log data from participants using each mode, as well as subjective feedback. The results provided valuable insight into the drawbacks and costs associated with each mode, leading to iterative refinement of the ML approach. Our institutional research ethics board approved the study (FILE 15755).

\subsection{Methodology}

\subsubsection*{Participants:} The criteria for eligibility within the study consisted of having some form of experience with digital pens in the past and owning a pen-enabled Windows-based device (to allow remote deployment during the Covid-19 pandemic). Seven participants used a digital stylus with a dedicated eraser, and eight used a device with barrel buttons. The remaining three used an onscreen button to enter the flick gesture mode. In total, 18 participants aged 18--39 participated (11 male/7 female). 16 participants were right-handed and 2 were left. We did not restrict the participation after pilot testing with left-handed lab members who achieved similar performance from both ML and Rule methods. 20 participants agreed to do the study, but complete data was only received from 18. Snowball recruitment was through university email lists to students, faculty, and staff, with an open invitation for participants to invite others. Compensation was given in the form of \$20 gift cards.

\begin{figure*}[t]
    \includegraphics[width=\textwidth]{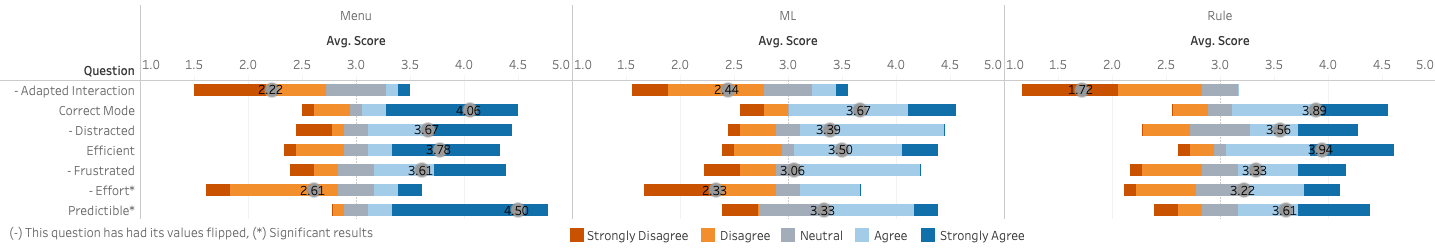}
    \caption{Subjective feedback --- significant differences were found for the level of Effort (Rule requiring less effort than ML, p<.05) and Predictability, with Menu better than ML and Rule, p<.05. Note some scales are reversed as indicated such that higher scores are always better.}
    \vspace{-1mm}
       \label{fig:likertmoderesponses}
\end{figure*}

\subsubsection*{Setup:} The study was conducted through a website that participants visited on their device on their own time (without synchronous assistance). The study consisted of a linear sequence of screens in which questionnaires and tasks were administered. Participants were asked to complete the study in one sitting of approximately one hour. Several rounds of pilot testing were used to refine task instructions, which were presented in text and video. Due to the study being conducted remotely, it was impossible to control the device's orientation. Instead, the device setup was collected, along with demographic information, in an initial questionnaire. All study screens are provided in supplemental materials.

\subsubsection*{Tasks:} The study consisted of a series of annotation tasks in a similar format to the initial data collection study. We varied the mode-switching technique across three conditions. Each condition was assigned a colour-coded term so the names would not bias participants. The three conditions (techniques) were: Red Mode = ML-based approach, Blue Mode = Rule-based approach, Green Mode = traditional toolbar menu. The order of conditions was counterbalanced.

Each of the conditions began with a \textbf{demonstration video}, followed by a \textbf{practice phase} in which participants performed at least five strokes in each of the three pen modes and used the flick gesture at least once. Directions were provided by marking the passages to highlight, underline, and write out in different colours. Passages were randomly selected to be about 15 words long. Participants were asked to continue until comfortable with the method. The second \textbf{markup phase} asked the participants to annotate a page of a text document, but this time they were asked to annotate all marked passages fully and correctly. Annotations could be made in any order, interleaving modes if desired. Strokes made in practice and markup phases were recorded to investigate learning strategies (practice) as well as performance (markup). After the markup phase, participants completed a feedback questionnaire consisting of seven 5-point Likert-scale questions and an opportunity for free text feedback. Likert questions consisted of statements in the form ``I felt <\textit{adjective}> with the <\textit{colour}> technique.'' (strongly disagree -- strongly agree, see Table \ref{tab:EvalStudy1Q}). Questions were designed to explore satisfaction with the technique, as well as hidden costs in terms of mental strain or complexity. Participants were also asked whether they used the mitigation technique (i.e., flick gesture), and if so, whether it was \textit{fast} and \textit{intuitive}. If it was not used, they were asked if they had any mistakes to fix, in order to gauge when people chose to erase rather than use the flick gesture.

After each condition, participants were offered the opportunity for a break. These steps were repeated for each condition for a total of 18 participants x 3 conditions x 2 phases per condition = 108 annotation tasks. After completing the tasks for all three conditions, a final questionnaire asked participants to rank the techniques from most to least preferred (tied ranks were not allowed). A text field was provided for any final comments on the approaches used.

%% file: 5b-comparisonResults.tex
\subsection{Results}

With the exception of one participant that took an extended break, participants took roughly four minutes to complete each phase of the study, which included marking up the provided document and answering questions.

\subsubsection*{Stroke Data}

Inspection of the strokes made within the study and modes revealed some interesting trends (Figure \ref{fig:strokeresults} (left)). In this figure stroke modes are colored based on their final mode (if corrections were applied). When using the Menu technique, participants would commonly annotate the provided document in a block manner, marking up all sections that required one pen mode before moving to the next. This is particularly visible with P9,13,14,16,17,18. For example, in the Menu technique, P13 completed all highlights, then underlines, then draw strokes. However, this pattern was not common with either of the automatic methods. Overall, 29.3\% of strokes in the Machine-learning technique used a different mode than the previous stroke, compared to 22.2\% in Rule, and 10.6\% in Menu. The block sequence of modes in the Menu technique may be a result of all stimuli being revealed at once (that is, in a real annotation technique, one would not know in advance all passages to highlight vs. underline). However, this result shows the different workflows adopted by participants based on the characteristics of the tools (automatic methods encouraged more mode-switching).

\subsubsection*{Corrections and Erasures}

Corrected strokes using the flick gesture indicate a mode failure, either automatic misclassification or a failure to change the mode with the Menu. Erased strokes may represent mode errors or participant mistakes when writing. In total, 749 strokes were edited in some way (15.2\% of all strokes made in the markup phases). Of those, 212 strokes (28.3\% of edited strokes) were mode-corrected with the flick gesture and 537 strokes (71.7\% of edited strokes) were erased (see Figure \ref{fig:strokeresults} (right)). The fraction of strokes erased or mode-corrected increased from Menu to Rule to ML for both. The lower rate of mode corrections for Menu is likely due to the user explicitly choosing the mode, but interestingly the correction technique was still used occasionally. For the Rule method we hypothesize that the rate may be lower than ML because participants learned to adapt their behavior to the predictability of the rules. Participants provided comments to improve the flick gesture, with the two most common themes being concerns about activation accuracy and the need to activate it in open space. P5 said, \textit{``There were multiple times in which I attempted to switch the modes with the eraser flick, it would erase something that I had done previously, causing me to have to redo the lining or highlighting I had just done.''}

\subsubsection*{Subjective Feedback}

After each markup task, the participants were asked to rank each of the three mode-switching techniques. In total, 8 participants chose Rule as their favorite, with 7 preferring Menu and 3 choosing ML. In terms of the lowest-ranked techniques, Rule was chosen 4 times, 6 chose ML, and 8 chose the Menu as their least favorite. A Wilcoxon signed-rank test (p = .05) was performed on all the pairs of rankings with results determining that no specific technique was significantly preferred over another. The rankings show a split between the predictability of the Menu approach (7 favorites) versus the efficiency of one of the  automatic methods (11 favorites) --- hinting that any automation is a matter of personal preference. This polarization is supported by the fact that the Menu technique received only 3 ranks of `2'.

The responses to the Likert questions are summarized in Figure \ref{fig:likertmoderesponses}. Note that the scales for \textit{Adapted Interaction}, \textit{Distracted}, \textit{Frustrated}, and \textit{More Effort than Usual} have been reversed so that higher values are better for all rows. Measures were compared pairwise with a two-tailed Mann-Whitney U test at p<.05. The Menu technique was most predictable compared to ML (U=96, p=.038) and Rule (U=65, p=.002). The menu was also deemed most correct, but no significant differences were found. The Rule technique had the best score in terms of effort and was significantly lower than ML (U=99, p=.048). Also, participants noted that they strongly adapted their interaction to use the Rule technique, showing they may have learned the hidden rules to achieve correct classifications. The degree of adaptation was lower for the ML technique but did not reach significance.%

The results suggest that while neither automatic technique was significantly preferred over the menu, automated approaches combined were preferred by 11 participants. When testing the two mode-switching techniques, people favoured the Rule implementation overall. Participant comments point to misclassifications in ML being the main challenge. P4 said \textit{``Red (ML) was very similar to blue (Rule) for me but it seemed like it wasn't as intuitive as the blue mode.''} Specifically, short strokes seem to challenge the ML technique, with P8 saying \textit{``I found it difficult when writing horizontal lines in my letters like f and t''} and P15 and P16 mentioning the dots of i and j being misclassified, leading to the need to erase the whole word. Though, some participants thought the ML technique was best. P2 said, \textit{``The red (ML) technique felt the smoothest of them all.''} and P16 thought the automatic techniques were \textit{``very helpful and saves a lot of time''} and were essentially tied, saying \textit{``red (ML) and blue (Rule) methods were both intuitive and saved me time moving the pen across the screen to select the tools.''}

\begin{figure}[t]
    \includegraphics[width=\columnwidth]{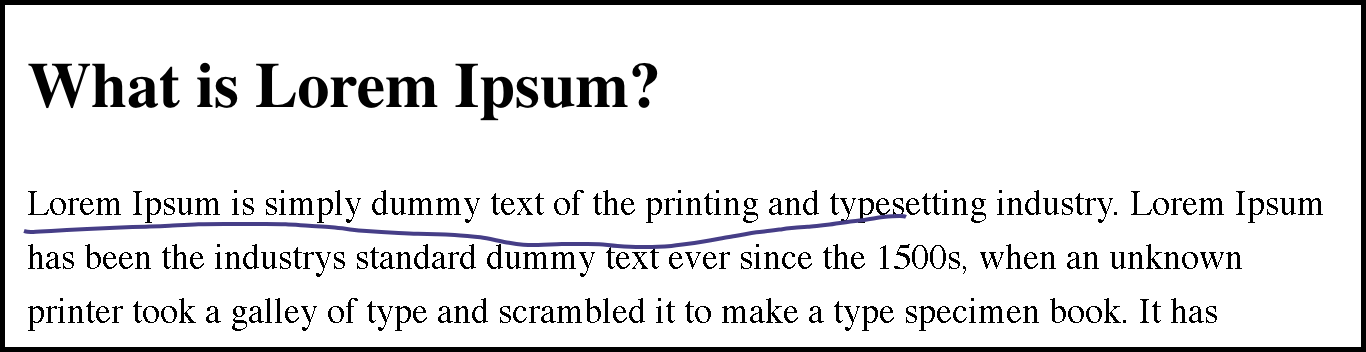}
    \caption[An Example of an Underline Stroke that has Drifted from its Intended Location]{An example underline stroke made by a participant.\@ While intended to be straight underneath the sentence, the stroke drifts towards the sentence below. }%
    \label{fig:UnderlineDrift}
    \centering
\end{figure}

\begin{figure*}[t]
    \includegraphics[width=\textwidth]{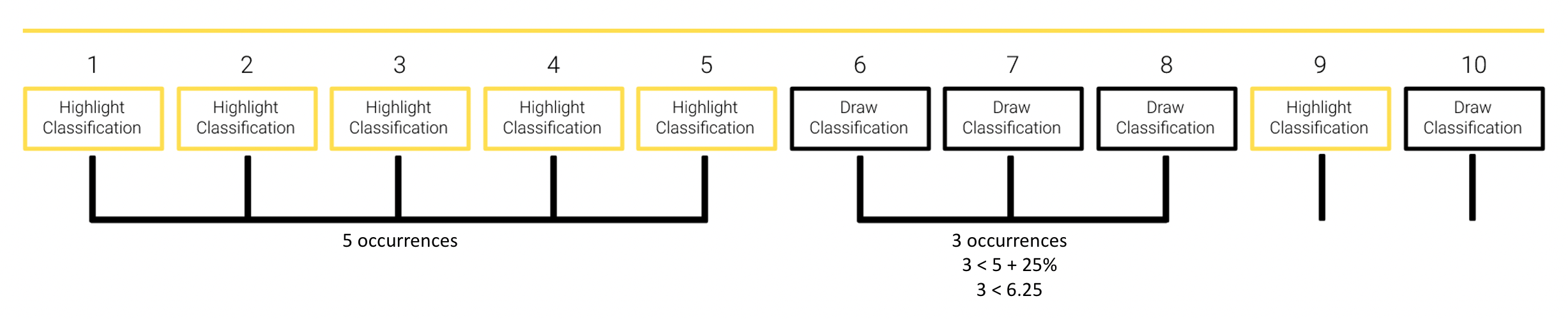}
    \caption[A Visual Demonstrating the Revised Mode Switching Technique]{The revised mode switching technique relies on a new class having N+25\% classifications before switching. In the example, an initial highlight mode switch occurred, while a subsequent draw mode switch did not occur because it did not collect 25\% more sequential classifications.}
    \label{fig:SeqModeSel}
    \centering
\end{figure*}

Regarding stroke corrections, the Rule technique had a higher concentration of errors made within the initial practice phase and had the largest drop when moving to the markup phase. We hypothesize that this stems from how comparatively simple it is to infer how the decision tree produces a stable result, allowing participants to adapt their interaction to better use its abilities. This claim is backed by the Likert responses from participants (Figure~\ref{fig:likertmoderesponses}), where they signified that they adjusted their interaction more for the Rule technique than any other. P18 said, \textit{``The Blue (Rule) technique simply had the best "logic" for handling mode switching.''}

In terms of comparing the Rule technique and the standard Menu approach, the decision seemed to be determined mainly by personal preference for automation vs. familiarity and predictability. P8 said, \textit{``I liked the Green (Menu) technique, since I had more control of the tool that I wanted to use.''} P15 said, \textit{``Green (Menu) technique was easier since I knew I did not have to expect or account for any mistakes other than my own.''} In contrast, others found the Menu overly inefficient compared to the Rule technique and even commented on the requirement of remembering what mode you're in while using it. P2, who did their menu work in mode blocks (see Figure~\ref{fig:strokeresults} (left)) preferred the menu but drew attention to how that preference would change if the work required frequent mode switches, saying \textit{``Would become very frustrating if frequently switching between pen modes.''}

This personal preference aspect suggests that an ideal interface should offer both options in terms of mode-switching techniques: an automatic method for frequent switches or those who prefer efficiency and a manual menu-based approach for infrequent switches or those who prefer predictability. A hybrid of both approaches may be best, as suggested by P13 who said \textit{``I believe a combination of Green (Menu) and Blue (Rule) would be perfect for me.''}

%% file: 5c-comparisonDiscussion.tex
\subsection{Discussion}
Our results revealed advantages and drawbacks for both Magic Pen approaches.  It is possible to reach 85.7\% accuracy on three-mode classification using an LSTM trained on only 9 participants. A manually created Rule-based approach reached 72.0\% accuracy. Despite the high accuracy of the ML approach on hold-out data, the Rule technique was preferred. These results point to several improvement opportunities for ML approaches. First, making techniques which are predictable seems essential to satisfaction. Also, any automatic approach has to maintain accuracy for short strokes as well as long ones.

There are some nuances to setup of our comparative study which may have disadvantaged ML more than Rule. Specifically, the LSTM model was trained in a fixed orientation, then deployed in a variety of device orientations. Due to the inclusion of pen pressure as a feature, device orientation may change the pressure profile, affecting classification accuracy, whereas the Rule-based approach is only reliant on ink position relative to onscreen elements, which does not change with device orientation.

Despite the challenges with misclassifications in both approaches, 11 participants preferred one of the automatic methods over the traditional menu-based approach. This shows receptivity to automatic approaches, and the promise of additional participant satisfaction with greater model or rule accuracy, something likely possible with more training data. While some participants enjoyed the flick gesture for mode correction, others complained that it incurred the cost of accidental erasure of strokes as it had to be used in a blank space which was not always convenient or possible.

Some participants preferred the Menu approach, as they did not like having a lack of confidence that the mode would be correctly selected. P15 summed it up, saying \textit{``[with Menu] you don't have to proceed with caution when writing or editing''}. This is fundamentally an issue of trust in the automated system and illustrates the impact of errors. Improved accuracy, predictability, and faster error mitigation may address these concerns.

The advantage of Magic Pen over Menu-based mode switching is also dependent on the task. Our study guided participants to the annotations to make, allowing them to plan their work. Mode planning was evident in the block structure of mode sequences in the Menu technique. Mackay previously showed that when workflows are planned, mode-switching can be reduced~\cite{mackay2002interaction}. In a more natural active reading scenario, akin to the \textit{problem-solving workflow} in Mackay's study, mode switching would be more frequent, increasing the burden of round-trips for the Menu technique. This calls for additional research to compare approaches in more natural annotation scenarios, where mode planning is not possible.

%% file: 6-post_study_improvements.tex
\section{Refining Magic Pen}

From the study findings, we focused on improving the accuracy and stability of the \acrshort{ml} approach. The Rule-based approach was abandoned due to lower accuracy, difficulty in manually curating the decision tree, and the influence it had on participant behaviour. Several factors of the \acrshort{ml} model, interface design, and mitigation technique were tweaked as a means to achieve overall improvements.

\subsection{Updated Mode Classification}

The results of the comparative study revealed that the most evident challenges of the \acrshort{ml} approach were (1) when the strokes were too short to confidently classify and (2) when strokes differed greatly from the norms the model had learned (see Figure \ref{fig:UnderlineDrift}). An example of the first type is the short stroke to dot an \textit{i} or \textit{j}. An example of the second type is when an underline stroke, typically located at the bottom of a line, drifted into the line below. This difference caused confusion between underlining and highlighting.

Heuristic methods were investigated to aid the model in resolving ambiguous strokes by relying on annotation-specific trends. The \textit{rapid heuristic} used the stroke class of the previously drawn stroke,  based on observing participants in the comparison study creating strokes in the same mode in rapid succession. This phenomenon was used to directly classify a subsequent stroke with the same pen mode if the time between strokes was less than a threshold, bypassing the \acrshort{ml} model. In testing, 500ms was optimal as it aided classifications for ambiguous, short strokes (e.g., the stick and dot in the letter `i').%

In this refined classification pipeline, when the next stroke is drawn outside the threshold window of 500ms, the previous stroke's mode still informs the classification, as follows. Through observation of the nature of errors of the \acrshort{ml} approach in the comparison study, we discovered that the model would produce a series of low-confidence classifications for some strokes and annotation styles. These low-confidence classifications would occasionally occur often enough in a row to incorrectly change the pen's mode. To avoid this, a \textit{confidence heuristic} is applied: if the model's confidence of the last window classification is below 75\%, the previous mode's stroke was used in place of the model's classification. 75\% was selected since any lower values were found to still allow for sub-optimal classifications and higher values began to block the use of new, correct classifications. Unlike the rapid heuristic-based technique, which uses the last stroke's pen mode, the confidence heuristic does not change the final mode of a stroke; only a single-window classification is replaced. All classifications must still be processed through the mode selection technique, which filters out erratic classifications from changing the final pen mode.

\subsection{Updated Mode Selection}
The mode-selection technique is critical to the implicit pen-mode detection pipeline because it is the step that, ideally, allows for correct mode changes while also filtering out short-term incorrect classifications. While the initial implementation selected modes based on whether the last $N=3$ classifications agreed, it was clear after the initial evaluation that it needed to be refined. The main issue was that, at lower majority threshold values, too many incorrect mode changes were allowed, even changing the mode near the end of drawing a long stroke. However, when the threshold was increased, there would be fewer mode changes, which affected correct changes.

\begin{figure}[t]
    \centering
    \includegraphics[width=0.8\columnwidth]{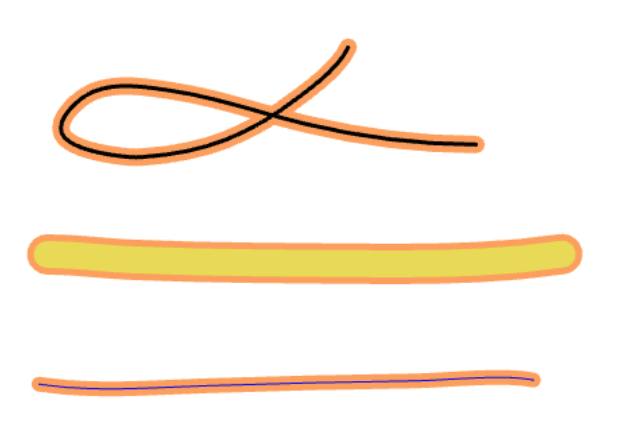}
    \caption{Examples of stroke highlights for draw, highlight, and underline modes. Strokes are briefly highlighted after they are completed and can be quickly removed by tapping the screen during this period.}
    \label{fig:StrokeFlagEx}
\end{figure}

\begin{figure*}[t]
    \centering
    \includegraphics[width=0.8\textwidth]{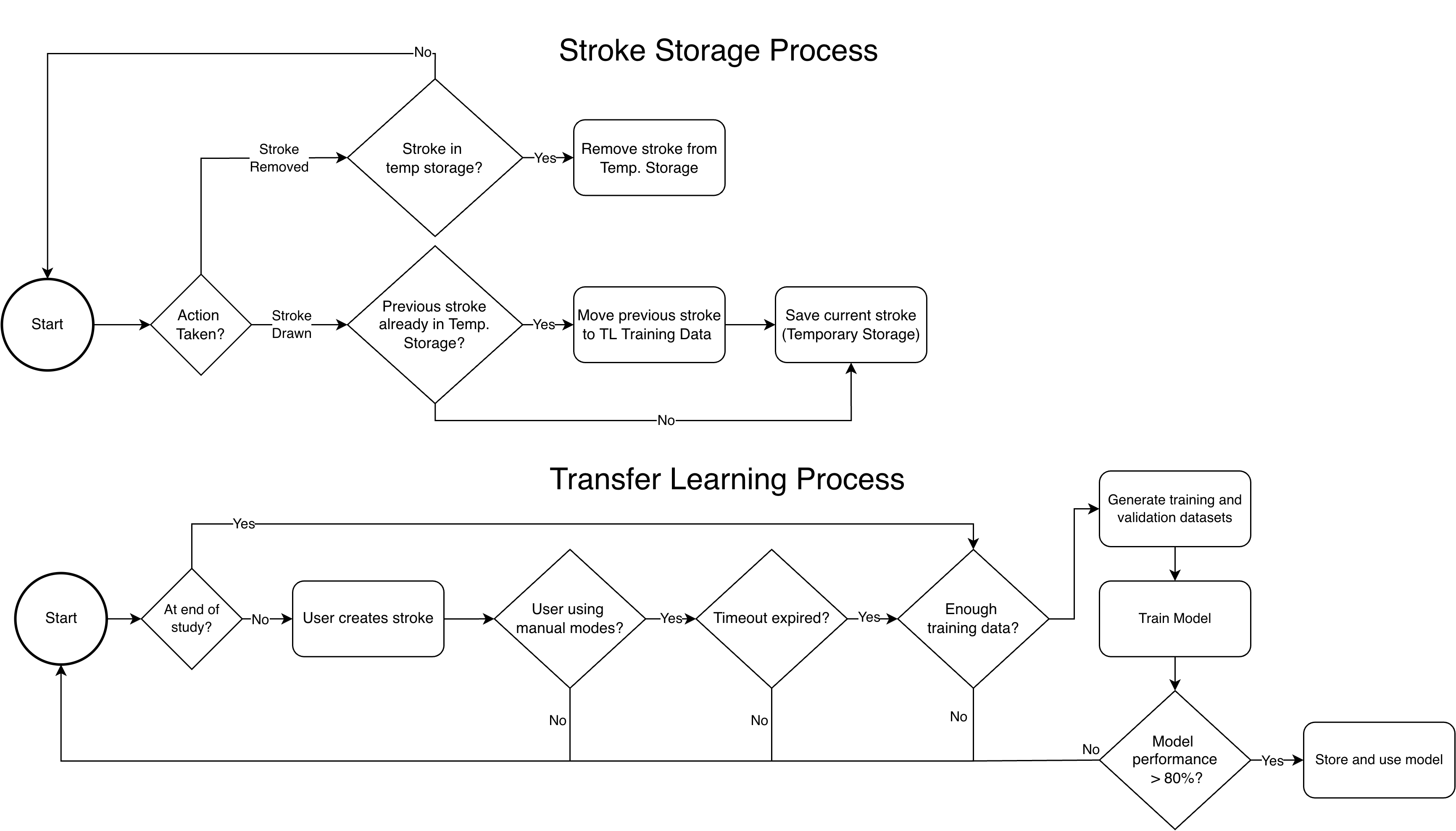}
    \caption{A flowchart visually outlining the process used for stroke data collection and \acrshort{tl} procedures.}
    \label{fig:tfflowchart}
\end{figure*}

The revised mode-selection technique chooses stroke modes based on the number of sequential occurrences of classifications (Figure \ref{fig:SeqModeSel}). For a new mode to be used, it must have 25\% more sequential occurrences than the current stroke mode. The main benefit to using this technique over the original is that it allows for a form of stroke-mode confidence to be built. If a series of new classifications arrive and surpass the existing amount, this indicates confidence in the new mode. In addition, it prevents a form of pen-mode thrashing (rapidly changing modes) which could confuse and annoy users.

\subsection{Updated Eraser Mitigation}

Although the previously implemented mitigation method for incorrect strokes aimed to avoid the frustration from making a stroke in the incorrect pen mode (due to the automatic mode-switching), it was not used often. Participants preferred to remove and redo their stroke rather than correct the mode. The primary reason was that the technique required its dedicated gesture for activation. Although participants were taught and practiced this gesture, it was evident that the extra work required to use it deterred people from using it. A source of cognitive load in the mitigation was the sequence of pen modes (see Section \ref{chap:EraserMitigation}) which were required to be memorized to be used effective. Another factor that inhibited the method's viability was the overloading of the method onto the pen's eraser tip. As participants noted, there were times when they wanted to change modes but would accidentally remove a stroke instead, requiring extra effort to undo the action or to re-draw it. While the threshold for determining a ``flick'' versus an ``erase'' could be tweaked, longer flicking windows would adversely affect the erasing interaction and vice versa. Therefore, for a novel mitigation method to be productive, it must have a clear activation method that can be distinguished.

\begin{table}
    \centering
    \small
    \setlength{\tabcolsep}{4pt}
    \begin{tabular}{ rrrr }
        \toprule
        Participant ID & Accuracy General & Accuracy TL & \% Change \\
        \midrule
        1 & 86.41 & 90.83 & 4.43\\
        2 & 85.95 & 89.26 & 3.32\\
        3 & 93.79 & 95.08 & 1.29\\
        4 & 90.37 & 94.88 & 4.51\\
        5 & 89.89 & 94.20 & 4.31\\
        6 & 74.76 & 85.62 & 10.86\\
        7 & 89.54 & 94.18 & 4.64\\
        8 & 89.10 & 92.01 & 2.91\\
        9 & 72.84 & 86.71 & 13.88\\
        10 & 94.16 & 94.83 & 0.67\\
        11 & 89.05 & 91.40 & 2.36\\
        12 & 87.00 & 87.67 & 0.67\\
        13 & 82.13 & 81.55 & -0.58\\
        14 & 94.96 & 95.90 & 0.94\\
        15 & 72.06 & 76.50 & 4.44\\
        16 & 94.18 & 97.02 & 2.84\\
        17 & 90.65 & 93.14 & 2.49\\
        18 & 94.55 & 96.52 & 1.97\\
        19 & 91.57 & 93.27 & 1.70\\
        20 & 78.20 & 91.60 & 13.41\\
        21 & 93.14 & 95.23 & 2.09\\
        22 & 96.30 & 96.75 & 0.45\\
        23 & 87.59 & 91.81 & 4.22\\
        24 & 68.82 & 70.18 & 1.36\\
        25 & 92.75 & 93.93 & 1.18\\
        26 & 94.25 & 96.03 & 1.78\\
        27 & 94.42 & 95.08 & 0.66\\
        \midrule
        &&Average:& 3.44\\
        &&Std:& 3.60\\
        \bottomrule
    \end{tabular}
    \caption{
        Hold-one-out experiment results for classification of participant data, before and after transfer learning. 20\% of the participant's data is used for TL, with the remaining 80\% used for evaluation purposes. The model is trained with all other participants' data before TL and evaluation. \label{tab:TFPerf}
    }
    \end{table}

Rather than mode switching, we turned our attention to facilitating quick stroke removal, which could be used for both mistaken strokes and mode errors. We experimented with different gestures, with the constraint of avoiding an erase mode button requiring a round trip to the toolbar.
Due to its prevalence in previous work~\cite{lank2006concurrent}, a tap on the screen was chosen to remove the most recent stroke. A tap is distinguishable from other touch-screen interactions (e.g., touch scrolling). To clarify the availability of stroke removal, a translucent orange highlight (see Figure \ref{fig:StrokeFlagEx}) is placed around the most recent stroke for 5s after the pen lifted, or until the next stroke is drawn.

%% file: 7-transferlearning.tex
\section{Transfer Learning}

After the evaluation study, it was clear that a method to tune the approach towards a user was needed to adapt classification and improve accuracy. Indeed, previous studies such as Marshall's landmark study of annotation~\cite{marshall1997annotation} observed that strokes are highly individual, implying that any system for so strokes made in a digital-form should respect this fluidity. To achieve this goal, \acrfull{tl} was leveraged to train an iteration of the model on one user's annotations --- effectively tuning it towards their interaction style.

\begin{figure}[t]
    \centering
    \includegraphics[width=\columnwidth]{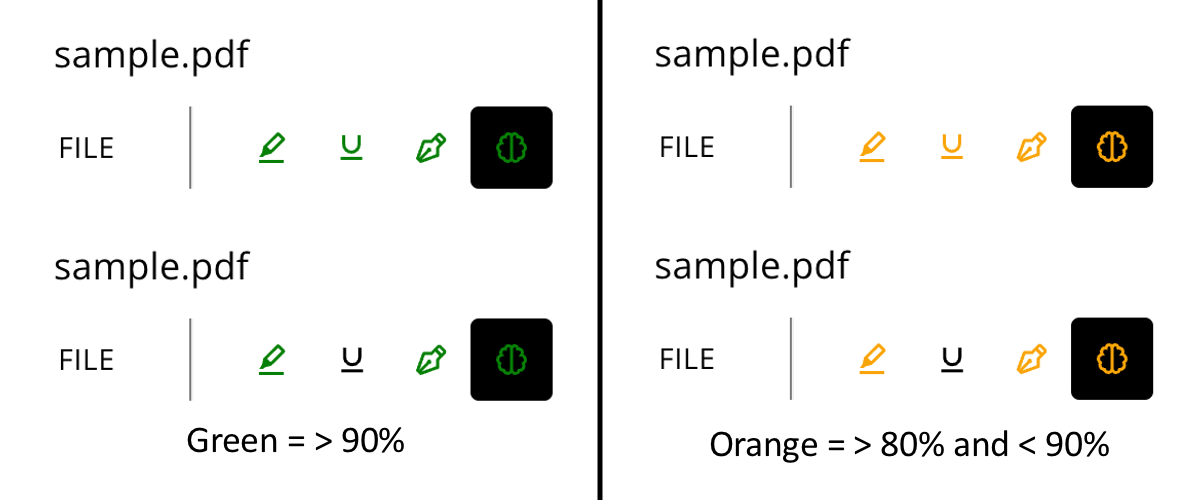}
    \caption{An example of the Magic Pen mode button changing colours based on model performance. In the top row, the model was trained on all three pen modes, while the bottom was only trained on two, signified by the colouring of the mode icons. Orange denotes performance between 80\% and 90\% while green denotes performance greater than 90\%. The automatic mode is selected in all cases.}
    \label{fig:ToolbarButtons}
\end{figure}

\acrshort{tl} is an \acrshort{ml} method in which a model is developed for a specific task, and then reused as the starting point for a second task. It is a popular approach within the field of deep learning due to its ability to allow for a pre-trained model to be used as a starting point, drastically reducing the compute and time resources normally needed~\cite{pratt1991direct}. \acrshort{tl} has been applied variety of tasks, such as cancer research \cite{khan2019novel, sevakula2018transfer}, text classification \cite{do2005transfer, semwal2018practitioners}, spam filtering \cite{tida2022universal, meng2010transfer}, and gaming \cite{zhu2020transfer}. A common factor for the inclusion of \acrshort{tl} is that it can increase the performance of a given network, while also not requiring a large dataset.

A common use of \acrshort{tl} is to take a pre-trained model, like those trained to detect cars, for example, and its knowledge learned for another task, like recognizing boats. However, this is not the only method in which \acrshort{tl} can be used. In Magic Pen, \acrshort{tl} is used to address idiosyncratic annotation styles, such as varying locations of underlines. This is backed up by prior research showing writing styles differ between people when using a traditional pen or digital stylus~\cite{schrapel2018pentelligence}.

\acrshort{tl} overcomes our small dataset problem but allowing for a pre-trained ``base'' model to be further trained with a small amount of individual user data, effectively tuning the iteration of the network towards that user. This customization also addresses generalizability challenges when pooling widely varying annotation styles.

\subsection{Performance}

A simulation was conducted to determine the potential of \acrshort{tl} to improve classification. Using the original model created in \acrlong{tf}, the set of annotation data collected from the Data Collection and Comparison studies was used as training data, while holding out one participant. Once training was completed, the model was tested on the holdout participants' data, followed by a \acrshort{tl} training sequence on 20\% of the holdout data. The \acrshort{tl} training sequence extracted the initial model's weights for each layer and used them within a new, participant-specific model as a basis. Both of the training sequences (initial training and \acrshort{tl}) followed the same training procedures as mentioned previously, with the \texttt{EarlyStopping} callback. The standard 80:20 ratio~\cite{gholamy201870} was reversed to 20\% training:80\% evaluation to mimic the extreme data limitation found within the annotation environment, where we want to build a custom model soon after a user begins to use the system. A learning rate of 0.0001 was used to reduce overfitting.

As shown in Table~\ref{tab:TFPerf}, across 27 participants, most were found to increase classification performance after \acrshort{tl} training. The highest recorded increase was from P9, from 72.84\% accuracy to 86.71\%, an improvement of 13.88\%. The average performance increase across participants was 3.44\%. We expect that, when implemented, the model will give continually improved performance with more data.

\subsection{Implementing Transfer Learning in the Annotation Interface}

With the performance benefits seen due to the introduction of \acrshort{tl}, the next step came to implement it within an annotation interface. To incorporate periodic \acrshort{tl}, we needed to answer three questions: (1) When do we collect stroke data? (2) When do we train on the stroke data? (3) When do we propagate model updates?  To solve the data collection problem, we could not just collect every stroke, as this could include incorrectly classified strokes, reinforcing classification errors. An assumption, based on observations from the comparison study, was made to solve this issue: if a stroke is in the incorrect mode, a user will remove it before making another stroke. Thus, all strokes which are not removed before writing another stroke are added to future training (see Figure~\ref{fig:tfflowchart} (top)).

To address the second question, on training, an always-learning model would be ideal. However, this was impractical as classifications could not be made while the model is learning due to performance constraints in the browser. To combat this challenge, background training with new stroke information was implemented at the end of a session or when the user switched to a manual pen mode. System performance could be maintained in manual pen modes while model training occurred in the background (see Figure \ref{fig:tfflowchart}). A timeout of 10 seconds after switching to manual modes before \acrshort{tl} training begins was implemented to prevent blocking the automatic method during active use of it. In addition, a training iteration is not triggered until a minimum of 500 stroke windows (roughly 10-20 strokes) are recorded. Each training iteration adapts the latest model with new strokes collected since the last training, incrementally building the model with new and recent data.

The third and final challenge is determining when to propagate (or keep) model updates and how to surface model performance to end users. To prevent the user from using a low-quality model, new model iterations are only kept (and made available) if their performance is greater than 80\% on a validation dataset of the user's own stoke samples. Because model re-training may produce varying levels of improvement, a method to inform the user of the model's performance was needed. To achieve simple quality visibility, the colour of the automatic mode method's button on the toolbar changes from grey (unavailable) to orange (performance between 80\% and 90\%) to green (performance greater than 90\%) (see Figure \ref{fig:ToolbarButtons}). Additionally, mode-specific accuracies are shown when the automatic mode is selected, by changing the toolbar icon colors for each mode.
\begin{table*}[t]
    \centering
        \begin{tabular}{@{} rcl }
         \toprule
         \textbf{\#} & \textbf{Session} & \textbf{Question} \\
         \midrule
         1 & All & \textbf{Accurate: }The system predicted my actions correctly.  \\
         2 & All & \textbf{Corrected:} I had to correct the system frequently. \\
         3 & All & \textbf{Predictable:} Results from the system were expected or predictable. \\
         4 & All & \textbf{Efficient:} The system was efficient to use. \\
         5 & All & What environment did you use the tool/system in (e.g. at a table, on a couch)? \\
         6 & All & Were there any modes that worked better or worse than others? \\
         \midrule
         8 & Final Session & \textbf{System Learned: }The system improved itself over time. \\
         9 & Final Session & \textbf{Distracted: }I found myself more distracted when using the system compared to a toolbar. \\
         10 & Final Session & \textbf{Frustrated: }I found myself frequently frustrated with the system. \\
         11 & Final Session & \textbf{Effort: }The system required more effort than a standard toolbar. \\
         12 & Final Session & \textbf{Preference: }I would use this system to annotate documents. \\
         \bottomrule
    \end{tabular}
    \caption{
        \label{tab:EvalStudy2Q}The questions for the deployment study feedback questionnaire. All questions except 5--6 were on a 5 point Likert scale.
    }
\end{table*}

%% file: 8-webannotationsystem.tex
\section{Web Annotation System}

\begin{figure*}[t]
    \centering
    \includegraphics[width=0.8\textwidth]{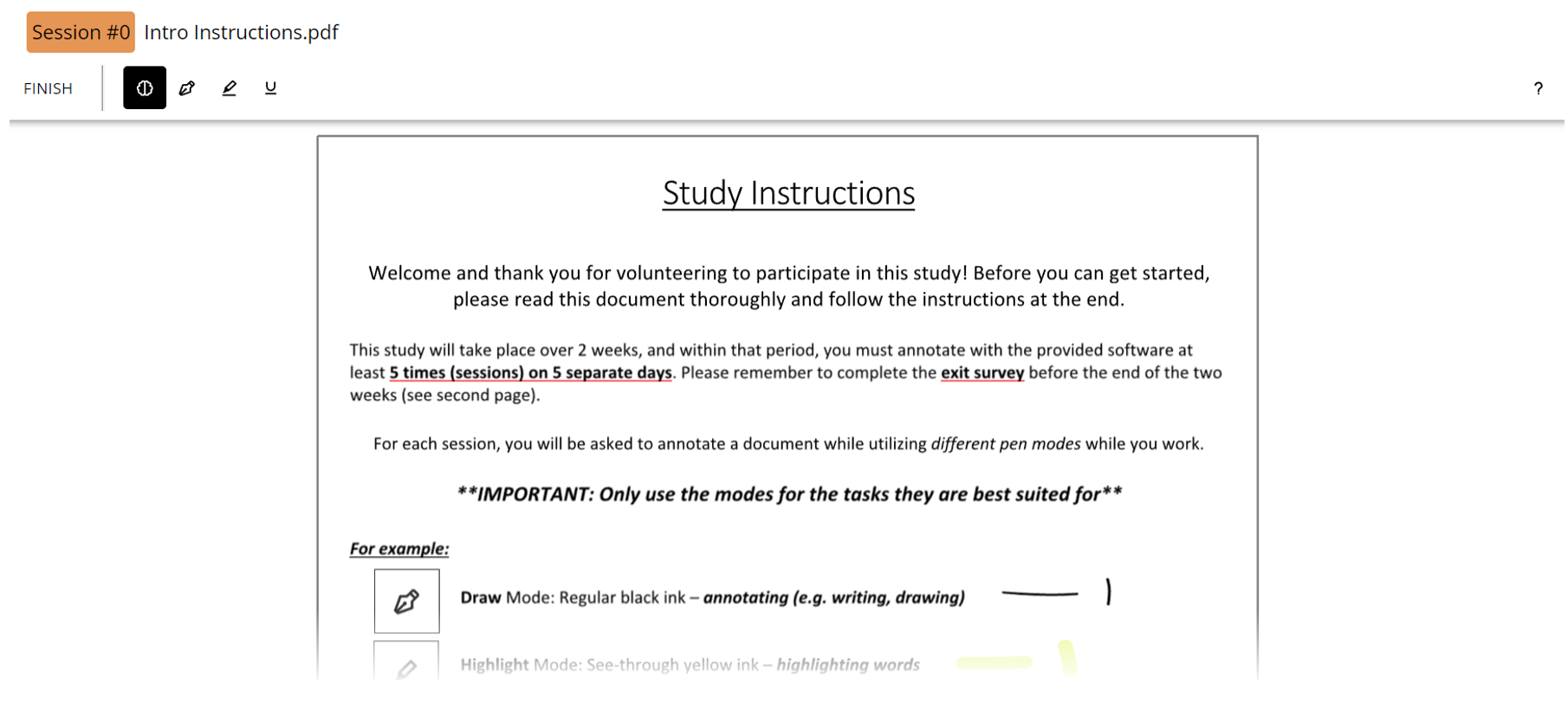}
    \caption{The deployment study interface, with the initial tutorial document loaded (cropped).}
    \label{fig:deployment}
\end{figure*}

\begin{figure*}[t]
    \centering
    \includegraphics[width=\textwidth]{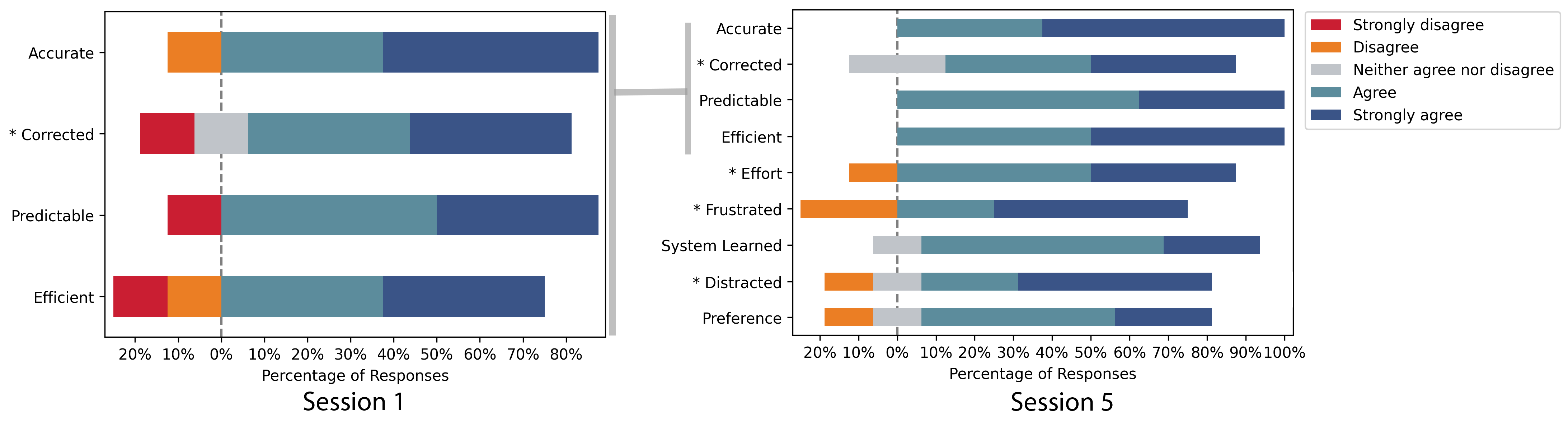}
    \caption{The subjective feedback Likert-scale responses from the deployment study, first session and final session. All measures improved over the sessions. The extra questions in the final session related to the overall experience. * indicates a reversed scale so that higher scores are always better.}
    \label{fig:deploymentLikert}
\end{figure*}

In order to deploy the final system in a study, we embedded the improved classification pipeline, including \acrshort{tl}, into a web-based annotation system.  To support the goal of allowing for naturalistic annotation, the system supports uploading of PDF documents, annotation, and downloading of the annotated documents. The MuPDF document viewer was used to convert PDFs to SVG and back\footnote{https://mupdf.com/}. A toolbar (Figure~\ref{fig:ToolbarButtons}) was provided to allow for manual or automatic mode selection. The system runs on Chromium-based browsers with pens that support W3C pointer events~\cite{w3c-pointer}.

The classification pipeline was moved from the server-side Tensorflow module used in the comparison study to a client-side implementation using Tensorflow.js (TFjs)~\cite{smilkov2019tensorflow}. This removed the network latency of communicating strokes and classifications and supported model personalization on-device. In a deployed (non-study) application, this architecture would better preserve user privacy by keeping documents and annotations fully local. Web Workers were used to offload model training and classification from the UI CPU thread~\cite{green2012web}, maintaining application responsiveness.

SVG rendering performance was slow for long documents or when many annotations were made. To alleviate this, we used the PaperJS library to render documents using the HTML5 Canvas\footnote{http://paperjs.org/}. On screen text box locations were cached to speed up the calculation of pen position features with respect to text. PaperJS also supports converting the entire canvas, including annotations, back to SVG, which is then converted to PDF using PDFKit\footnote{https://pdfkit.org/} for download. For the deployment study, redacted PDFs are also created during this conversion and submitted for evaluation.

%% file: 9-deploymentStudy.tex
\section{Deployment Evaluation Study}

To measure the impact that \acrshort{tl} has had on automatic annotation, and to determine if it is preferred to a standard toolbar, a final deployment study was conducted. The study collected log data, similar to the comparison study, and was approved by our institutional research ethics board.

\subsection{Methodology}
\subsubsection*{Participants:}
Participants were screened to have previous experience with digital pens and to own a Windows-based device to use for the study. Self-declared participant experience ranged from novice to expert. All participants used a digital pen with a dedicated eraser end. In total, 8 participants aged 18--29 (5 male / 3 female) were recruited, all right-handed. Recruitment was done through university email to students, faculty, and staff, with an open invitation for students to invite others. Although limited to new participants, there was potential for previous study participants to contribute, as identifying information for previous participants had been removed per REB guidelines. Compensation of \$10 was given per measurement session, up to \$50 for completing all 5 sessions. All participants completed the introduction + 5 sessions.

\subsubsection*{Setup:}
The study was conducted through a website that participants visited on their own devices on their own time. Multiple sessions were used over a two-week period in which the user could access the site any time they wished, as long as a minimum of 5 sessions were completed. Each session consisted of annotating a document, followed by completing an experience questionnaire. Participants were asked to complete a session in one sitting. Due to the study being also conducted remotely, device orientation was  impossible to control. Instead, the participants were asked to keep the device in a comfortable position and to try to remain in that position for the duration of the study.

\begin{figure*}[t]
    \centering
    \includegraphics[width=\textwidth]{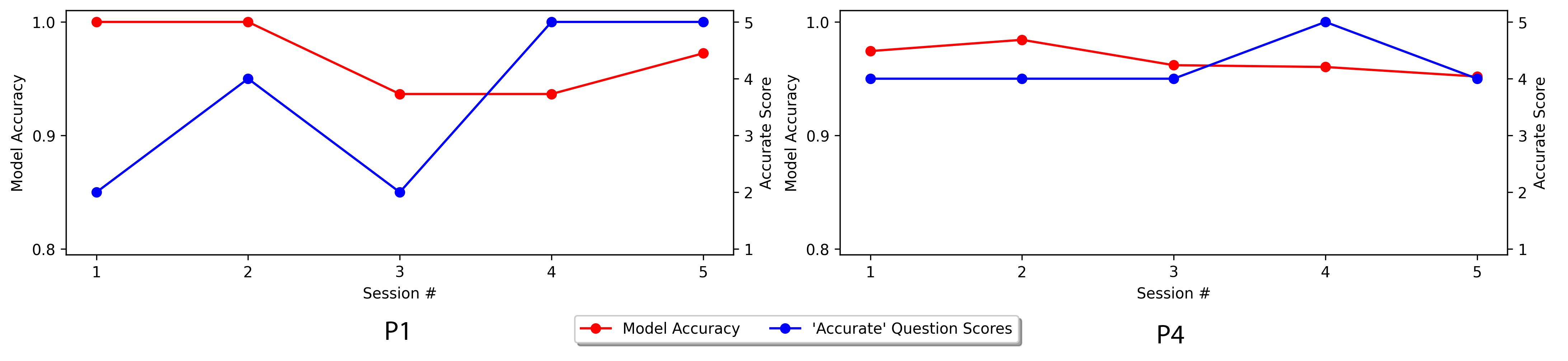}
    \caption{Perceived accuracy compared to model validation accuracy across sessions, for P1 and P4. Model validation accuracy was measured based on hold-out data (stroke windows) from the most recent training iteration.}
    \label{fig:accuracyComparison}
\end{figure*}

Sample documents were created by injecting errors (common misspellings, random capitalization, extra spaces) into 2.5\% of words of a supplied document. In pilot testing, this approach maintained readability while creating clear targets for annotation. Five relatively error-free high school student essays were selected and passed through the script to generate five error-prone documents. These essays were made available to participants.

The system collected strokes drawn, their time, and mode. Erased strokes were flagged but retained in the data. Model parameters and the accuracy on hold-out validation data was collected for each training iteration. A redacted view of the document and annotations was also submitted with the questionnaire responses at the end of each session.

\subsubsection*{Tasks:}
The study was structured to consist of similar annotation tasks from the comparison study. However, the goal of the deployment study was to collect more natural annotations and observe performance and preference over a longer deployment. Thus, participants were invited to upload their own \acrshort{pdf}-documents and to annotate normally while using the automatic and manual modes (which trained the automatic method) in any order. If participants did not have documents to annotate, they could use one of the supplied samples. To reduce the mode-block nature induced by the setup of the comparison study, sample document instructions only describe the type of errors that will be found and ask the user to correct them using whichever method they want, rather than instructing which pen modes to use and where.

The study began with an introductory training session, consisting of annotating a tutorial document. The tutorial document explained the automatic pen system, how to train the system, the eraser technique, and how many sessions to complete (see Figure~\ref{fig:deployment}). For the introduction session, participants were presented with an un-tuned version of the Magic Pen model, with specific annotation instructions embedded in the tutorial document to ensure the model could be trained at least once. Every subsequent annotation session began with the user uploading a \acrshort{pdf} document (their own or a provided sample). From there, participants would annotate as they saw fit, with the requirement to make at least 10 unique, somewhat long, strokes so that the model could be further trained.  When the participant was finished annotating, they pressed the "Finish" button. If not enough strokes were collected to retrain the model, a warning popup asked the participant to write a bit more, then try again. Otherwise, a final \acrshort{tl} iteration began, and the session questionnaire was presented (see Table~\ref{tab:EvalStudy2Q}). After the questionnaire was submitted, the model, data, and redated document were uploaded automatically to our study server. The participant was offered the opportunity to download a copy of the annotated file for their own use.

These steps, aside from the introductory session, were repeated for each session throughout the two-week period for a total of 8 participants x 5 sessions = 40 annotation sessions. At the end of the five sessions, participants were given the option of completing an exit survey to complete their participation in the study. Participants were allowed to continue to use the system to annotate as many times as they desired within the two-week period, with the only restriction that they complete the exit survey during their final session.
\begin{figure*}[t]
    \centering
    \includegraphics[width=\textwidth]{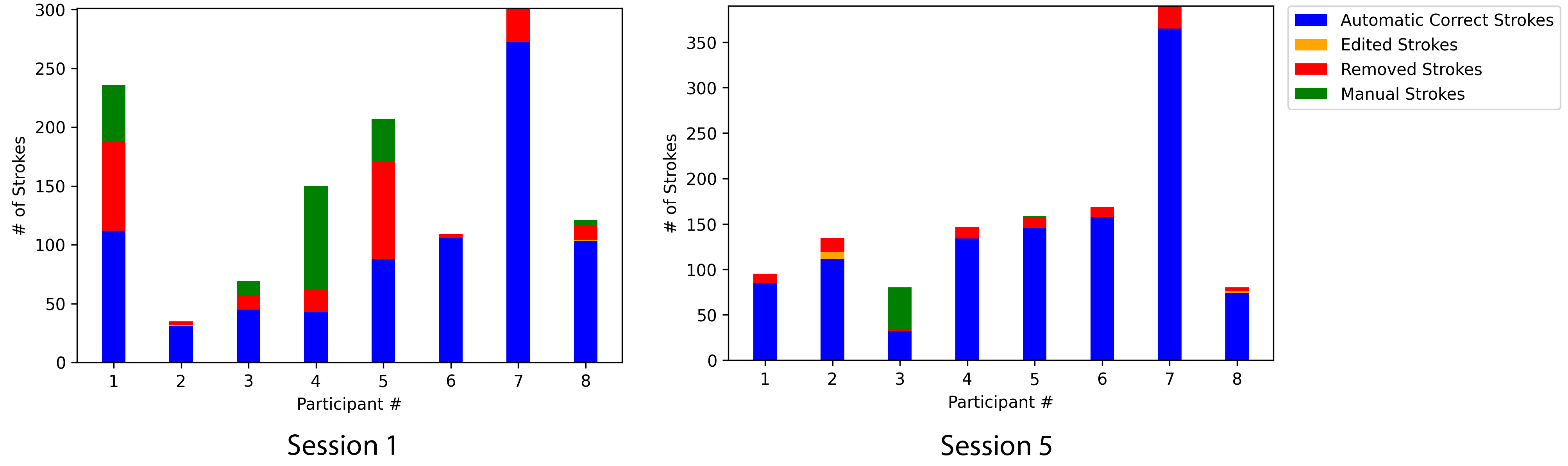}
    \caption{Stroke distribution by type and participant, across session 1 and session 5. The use of manual modes in session 5 was the same or lower than in session 1 for all but P3. Fewer strokes were removed in later sessions.}
    \label{fig:strokeModes}
    \vspace{-3mm}
\end{figure*}

%% file: 9b-deploymentResults.tex
\subsection{Results}
All participants completed 5 sessions, each of which took roughly 12 minutes on average. However, data from session 4 for P5 was lost due to a browser error.  Of all the participants, only one provided their own documents, with the rest using the provided samples.

\subsubsection*{Introductory Session}

The first session (session 0) acted as a tutorial for participants. Participants were instructed to annotate in a specific method similar to the comparison study, with an initial version of the \acrshort{ml} model, which was not customized to their writing style. Overall, all participants found the base model accurate (M=4.38, SD=0.48 on a 0--5 scale where 0 is `strongly disagree' and 5 is `strongly agree'). The participants also felt that the model was efficient (M=4.50, SD=0.71) and predictable (M=4.50, SD=0.71). While not strictly comparable due to the differing setup, it appears impressions of the base model were better than during the earlier comparison study: 7 participants indicated that they did not need to correct their strokes post-creation (M=4.38, SD=0.70).

Participant questionnaire scores were evaluated over the various sessions to determine the impact of \acrshort{tl} on the annotation experience, aside from session 0, since it used the base model. A summary for the first and final measurement sessions is shown in Figure~\ref{fig:deploymentLikert}. In the summary, all measurements improved between the first and final session.

When broken down by participant, there is more variation in the results. On the question of \textbf{accuracy}, the ranking results show either a net positive or no change in between the first and final sessions for all participants except P6. Each participant had at least one session in which perceived accuracy decreased, but the overall trend was positive across sessions. P6 had an overall decrease in both perceived and model validation accuracy, likely resulting from not providing enough training data through using the manual modes. P6 used the automatic mode exclusively, relying on removing erroneous strokes to provide model feedback.

The perception of accuracy suggests that the model improved overall performance, but the raw performance information of the \acrshort{tl} reveals a more complex story. Each time the model retrains, it records an accuracy score on a hold-out validation dataset of stroke windows from the participant's own recent strokes. We expected this score to be monotonically increasing across sessions, but it was not. Indeed, the accuracy varied with each retraining, between 80--100\%. Recall that 80\% was the cut-off for accepting a new model after retraining. The variation stems from the choice to use recent held-out stroke windows as the validation set for new model iterations. In retrospect, this makes accuracy comparison across models difficult to interpret, since the test data has changed. In addition, due to the unconstrained nature of the annotation task in the deployment, we saw high variation in annotation behavior in relatively short sessions, further resulting in variation in model accuracy. Examining the participants' annotated \acrshort{pdf} documents revealed that for the majority of training iterations, participants only made a single type of stroke with the manual method before the training cycle. We discuss the implications of this in the next section. An example of model accuracy on validation data compared to perceived accuracy, for two participants, is shown in Figure~\ref{fig:accuracyComparison}. A deeper investigation of the submitted logs revealed that for P1 in Session 1, the main cause of errors was the use of a document with tight line spacing, making it difficult to differentiate underline and highlight. Similarly, for P1, Session 3, the document used did not contain text elements, so the model could not pick up on word-level features. By the final session, the validation accuracy of the models decreased 5\% on average compared to their first \acrshort{tl} session. Despite the model's raw \acrshort{tl} performance, all but one participant ranked final accuracy either a 4 or 5, showing that it did not affect participants' satisfaction with the accuracy. Detailed results are in the supplemental material.

Inspection of the uses of the manual methods revealed some interesting trends amongst participants. Participants 1, 3, and 5 consistently made extensive efforts to invoke the model training through providing significant strokes in the manual modes, until the last session, when only P3 continued to use manual modes significantly. Averaged all sessions, all participants made more than 50\% of their strokes using the automatic mode, with increasing use over time. An example of the breakdown of manual, automatic, and removed strokes for sessions 1 and 5 appears in Figure~\ref{fig:strokeModes}.

There was a commonality amongst participants in reported mode errors, with the underline mode being reported by 7 out of the 8 participants. 7 of 8 participants also mentioned how well Draw/Highlight worked at the end of the first session. Reported mode difficulties for the model diminished as the participants continued through the sessions, with seven participants mentioning that all the modes worked similarly by the second-last session. This drop in mistakes can be seen in the number of stroke corrections needed over the sessions by participants in Figure~\ref{fig:strokeModes}. Upon detailed analysis of the log files and submitted PDFs, the high rate of removal for P1 in Session 1 was due to the underline mode errors discussed above. For P5, most of the removed strokes appeared to be correctly moded, and removed in a single erase event, suggesting that the participant removed them for other reasons. We suggest in future studies of this type, the number of \textit{immediately removed} strokes be counted differently from those removed at any later point in the session, as immediate removal is more likely to indicate a mode error.

\subsubsection*{Final Session Preferences}
In the additional questions from the final questionnaire (Figure~\ref{fig:deploymentLikert} right, bottom 5 rows), the results were generally positive. The Magic Pen technique was preferred among participants (M=3.88, SD=1.0). The few participants reporting lower scores for Effort, Frustrated, and Distracted were largely due to P1 and P3, who also did not prefer using Magic Pen. We compared the qualitative metrics to determine a relation to the lower preference metrics by computing the Pearson Correlation Coefficient. The results showed that Frustration was the highest correlate to Preference (R=0.78, p=0.02). This aligns with the \textit{Frustration} scores for P1 and P3. Despite feeling frustrated, both participants noted that the system actively learned throughout the study and was accurate by the final session. These results demonstrate that overall technique preference, while ultimately dependent on model performance, relates to more than just accuracy.

%% file: 9c-deploymentDiscussion.tex
\subsection{Discussion}

The deployment study has revealed several actionable insights which should be considered in the creation of future pen-mode inferencing systems like Magic Pen:

\textbf{Preference Relates to More than Accuracy:} The Magic Pen deployment overall preference results align with the comparison study's conclusion about the importance of both accuracy and predictability of algorithmic behavior. While most participants preferred the Magic Pen automatic mode inferencing, those who did not expressed frustration at the perceived unpredictable nature of the classifier. The predictability metric is thus a good supplemental measure of model improvement. This claim is backed by responses from participants, which point to an increased understanding of the reasons for mode errors with greater experience. For example, in response to the question, ``Were there any modes that worked better or worse than others?'' P1 said in Session 1, \textit{``Underlining seemed to be less accurate''}. In Session 3,  \textit{``Highlight was right the most, while I felt like I had to correct underlining the majority of the times I tried to use it''}. By session 4, P1 expressed \textit{``All worked well, underline and highlight sometimes got confused, but \textbf{that was usually my fault}''} (emphasis ours). Finally, in Session 5, they said, \textit{``No, all worked well''}. When the participant can discern how the model works, they can both adapt to improve accuracy, and reduce blame and frustration. While this goes against our original goal of a fully naturalistic mode inferencing system, perhaps some level of co-evolution of human and model behavior is inevitable and would result in preferred and powerful systems.

\textbf{Validate Against Long-Term Data:} While the accuracy varied up and down across sessions of \acrshort{tl}, this was likely an artifact of our measurement method. Our approach validated the new model  against a hold-out sample of recent stroke windows (i.e., since the last training). However, the nature of \acrshort{tl} is that the model is learning both long and short-term patterns.  We thus recommend that future iterative approaches to building custom mode inferencing models use a longer-term validation set. This could be built up by adding stroke windows with each subsequent training session. In this way, both the model and validation would be anchored in data across multiple sessions, which could increase the stability of model accuracy measurements.

\textbf{Consider Document Variations:} Unsatisfactory model performance for one participant was due to unforeseen aspects of the documents used (close line spacing, no text). This points to the need for training across not only a variety of people, but also a wide variety of documents, since some data features are tied to document elements. Additionally, heuristic approaches could be used to detect documents which do not match the characteristics of those in the training pool, and automatic modes could be disabled, or a warning presented to the user.

%% file: 10-conclusion.tex
\section{Conclusion and Future Work}

In this paper, we developed and investigated Magic Pen, a technique for low-latency mode classification based on implicit signals from the way people hold and use digital pens to annotate text documents. Simulations after an  initial data collection study demonstrated the feasibility of the approach. We compared two different back-end implementations of automatic mode switching, settling on a machine learning approach, while learning about the importance of predictability from the rule-based approach. We improved the machine learning approach with refined features, heuristic methods for classifying short and ambiguous strokes, and a simplified error mitigation technique. The next generation of the technique was augmented with a personalization step through iterative transfer learning, allowing the model to adapt to idiosyncratic styles of annotation and device orientations over time. Magic Pen reduces round trips to modal buttons and encourages fluid interleaving of modes.

The results suggest that participants overall preferred to use the refined automatic technique over traditional menu-based mode switching and that preference related to model accuracy, predictability, and the level of frustration.  Although there is room for improvement of Magic Pen, including expansion to other modes of annotation and a variety of documents, it is clear there is promise in implicit mode switching for annotation systems.

While our approach worked across many devices with low latency, significant engineering efforts were expended to make it work in the browser. With additional computing resources, such as the incorporation of GPU processing, future models could incorporate additional pre-stroke data features such as pen hover, or even non-stroke features such as gaze, which may improve classification accuracy and reduce latency even further.  With computing power, it also may be possible to implement continuous learning during active annotation, which would reduce the need to switch to manual modes to retrain the model. Machine learning approaches beyond \acrshort{lstm} could also be explored.

 There is also an opportunity to explore the effect of model understandability on performance. In the comparison study, some participants preferred to use a more easily understandable technique. Explainable ML techniques which break down the `black box' of how systems work may better support users to build a mental model of how the technique works and support co-evolution of the model and the annotation behavior toward an optimal state.

A final avenue for improvement is the opportunity to better infer user pen task intents beyond ink annotation modes. For example, could it be possible to automatically interleave selection techniques with inking modes of a digital pen, without explicit modes? Or to support concrete tasks such as queries for a definition of a word using the pen, without accidentally leaving ink on the page, or having to use dedicated buttons or learned gestures.

Given the success of our approach with a small feature set, in-browser classification, and training data initialized with just 9 participants, this work demonstrates the great promise of future predictive mode switching techniques to be directly integrated into the tools we use every day. As with any ML-driven approach, the important question of how to handle failure gracefully, and learn from it, is an important topic for additional investigation.